\documentclass[12pt]{amsart}
\usepackage{graphicx}
\usepackage{grffile}
\usepackage{latex_base}
\usepackage{macros}
\usepackage{pgfkeys}
\usepackage{subcaption}
\usepackage[super, comma, compress]{natbib}
\makeatletter           
\renewcommand\@biblabel[1]{#1\hspace{1em}} 
\makeatother
\setcitestyle{super, open={}, close = {}, {--}, aysep={; }, yysep={,}, notesep={, }} 

\usetikzlibrary{graphs}

\usepackage{array}

\newcolumntype{+}{!{\vrule width 2pt}}

\newlength\savedwidth

\newcommand\thickhline{\noalign{\global\savedwidth\arrayrulewidth\global\arrayrulewidth 2pt}%
\hline
\noalign{\global\arrayrulewidth\savedwidth}}

\makeatletter
\renewcommand*\@maketitle{%
  \normalfont\normalsize
  \@adminfootnotes
  \@mkboth{\@nx\shortauthors}{\@nx\shorttitle}%
  \global\topskip42\p@\relax 
  \@settitle
  \ifx\@empty\authors \else \@setauthors \fi
    \ifx\@empty\thankses\else {%
    \def\par{\let\par\@par}\footnotesize{\centering \@setthanks\@@par}}%
  \fi
  \ifx\@empty\@date \else {\vskip 1em \vtop{\centering\large\@date\@@par}}\fi
  \ifx\@empty\@dedicatory
  \else
    \baselineskip18\p@
    \vtop{\centering{\footnotesize\itshape\@dedicatory\@@par}%
      \global\dimen@i\prevdepth}\prevdepth\dimen@i
  \fi
  \@setabstract
  \normalsize
  \if@titlepage
    \newpage
  \else
    \dimen@34\p@ \advance\dimen@-\baselineskip
    \vskip\dimen@\relax
  \fi
} 
\renewcommand*\@adminfootnotes{%
  \let\@makefnmark\relax  \let\@thefnmark\relax
  \ifx\@empty\@date\else \@footnotetext{\@setdate}\fi
  \ifx\@empty\@subjclass\else \@footnotetext{\@setsubjclass}\fi
  \ifx\@empty\@keywords\else \@footnotetext{\@setkeywords}\fi
}
\makeatother

\makeatletter
\newcommand{\addresseshere}{%
  \enddoc@text\let\enddoc@text\relax
}
\makeatother

\author[de Wolff]{Timo de Wolff$^{*\ 1,2}$}
\address{Dr. phil. nat. Timo de Wolff, Technical University of Braunschweig, Institute for Analysis and Algebra, Universit\"atsplatz 2, 38106 Braunschweig, Germany\medskip}
\email{t.de-wolff@tu-braunschweig.de}

\author[Pfl\"uger]{Dirk Pfl\"uger$^{*\ 1,3}$}
\address{Dr. rer. nat. Dirk Pflüger, University of Stuttgart, Institute for Parallel and Distributed Systems, Universitätsstr. 38, 70569 Stuttgart, Germany\medskip}
\email{Dirk.Pflueger@ipvs.uni-stuttgart.de}

\author[Rehme]{Michael Rehme$^3$}
\address{Michael Rehme MSc, University of Stuttgart, Institute for Parallel and Distributed Systems, Universitätsstr. 38, 70569 Stuttgart, Germany\medskip}
\email{Michael.Rehme@ipvs.uni-stuttgart.de}

\author[Heuer]{Janin Heuer$^2$}
\address{Janin Heuer MSc, Technical University of Braunschweig, Institute for Analysis and Algebra, Universit\"atsplatz 2, 38106 Braunschweig, Germany\medskip}
\email{janin.heuer@tu-braunschweig.de}

\author[Bittner]{Martin-Immanuel Bittner$^{1,4}$}
\address{Martin-Immanuel Bittner MD DPhil, Arctoris Ltd, 120E Olympic Avenue, Milton Park, Abingdon, Oxfordshire OX14 4SA, United Kingdom\medskip}
\email{martin-immanuel.bittner@arctoris.com}

\thanks{$^*$Contributed equally}
\thanks{$^1$Young Academy of the German National Academy of Sciences, Berlin, Germany}
\thanks{$^2$Institute for Analysis and Algebra, Technical University of Braunschweig, Germany}
\thanks{$^3$Institute for Parallel and Distributed Systems, University of Stuttgart, Germany}
\thanks{$^4$Arctoris, Oxford, UK}


\title[]{Evaluation of pool-based testing approaches to enable population-wide screening for COVID-19}

\begin{document}

\maketitle

\bigskip

\begin{center}
    Corresponding author \\
Martin-Immanuel Bittner, \emph{MD DPhil}\\
Young Academy of the German National Academy of Sciences\\
Jägerstr. 22\\
10117 Berlin\\
Germany\\
\url{martin-immanuel.bittner@arctoris.com} \\
\end{center}

\newpage

\section{Abstract}

\subsection*{Objective:} Rapid testing is paramount during a pandemic to prevent continued viral spread and excess morbidity and mortality. This study investigates whether testing strategies based on sample pooling can increase the speed and throughput of screening for SARS-CoV-2, especially in resource-limited settings.

\subsection*{Methods:} In a mathematical modelling approach conducted in May 2020, six different testing strategies were simulated based on key input parameters such as infection rate, test characteristics, population size, and testing capacity. The situations in five countries were simulated, reflecting a broad variety of population sizes and testing capacities. The primary study outcome measurements were time and number of tests required, number of cases identified, and number of false positives. 

\subsection*{Findings:} The performance of all tested methods depends on the input parameters, i.e.\ the specific circumstances of a screening campaign. To screen one tenth of each country's population at an infection rate of 1\%, realistic optimised testing strategies enable such a campaign to be completed in ca. 29 days in the US, 71 in the UK, 25 in Singapore, 17 in Italy, and 10 in Germany. This is ca.\ eight times faster compared to individual testing. When infection rates are lower, or when employing an optimal, yet more complex pooling method, the gains are more pronounced. Pool-based approaches also reduce the number of false positive diagnoses by a factor of up to 100. 

\subsection*{Conclusions:} The results of this study provide a rationale for adoption of pool-based testing strategies to increase speed and throughput of testing for SARS-CoV-2, hence saving time and resources compared with individual testing.

\section{Introduction}

Pandemics such as COVID-19 pose a significant public health threat, leading to morbidity, mortality, and rapid and significant strain on the health system.\cite{Ezekiel}
Faced with the global spread of a novel pathogen, the identification of cases and carriers and elucidation of transmission patterns is paramount.\cite{Kelly-Cirinoe001179, Thecriticalroleoflaboratorymedicineduringcoronavirusdisease2019COVID19andotherviraloutbreaks}
The ability to rapidly and reliably diagnose those infected is critical to 1) identify and control clusters of infection; 2) prepare the health system for the patient numbers to be expected; 3) deploy medical countermeasures in a targeted way; and 4) assess the effectiveness of any public health measures and adapt them accordingly.\cite{Pulia}

The speed of testing is critical, but is limited by supply of and access to diagnostic tests, logistical challenges, and shortages in qualified personnel and/or laboratory facilities that could perform the necessary tests.\cite{Petom1163, Burki}

In each of the above scenarios, maximising the number of people that can be tested in a given time is essential. Universal weekly testing has been proposed as the only viable exit strategy from lockdown.\cite{gottlieb2020, wsj,petolancet2020, Petom1163, safra, audio} One potential approach of increasing testing efficiency is pooling of different samples in one test. This is a well-validated method used for example in transfusion medicine for HIV testing and has recently been experimentally deployed for SARS-CoV-2 in a small-scale study in California.\cite{doi:10.1056/NEJMoa042291,10.1001/jama.2020.5445}

Using a simulation approach, this study aims to identify the most effective testing strategy by comparing six mathematical procedures for mass testing a given population for infection with SARS-CoV-2. The primary objective thereby is to identify as many cases as possible as quickly as possible with a given limited testing capacity. In other words, we aim to deploy the available tests as effectively as possible, increasing the number of identified cases per test (ICPT).

\section{Methods}

\label{sec:mathematical_background}

\subsection*{Assumptions}

We simulate a screening campaign aimed at an entire population assuming a range of different possible infection rates $ir$ (including unreported cases).

We assume a testing capacity of $c$ (e.g.\ PCR-based) tests per day, and that each test takes 5 hours to process in a clinical laboratory, resulting in a capacity of $c\frac{5}{24}$ tests in parallel. For reasons of consistency, sample logistics are not included as an input parameter.

In terms of test characteristics we assume a sensitivity of $p=0.99$ based on data reported by LabCorp to the FDA and an estimated false-positive rate of $q=0.01$.\cite{LabCorp} We assume the test characteristics to remain constant after pooling, based on a maximum pool size of $k=32$ which has been shown to provide reliable results for COVID-19.\cite{IsraelPaper,lohse20pooling}

To account for the case of a potential drop in sensitivity due to sample pooling via a dilution effect, we duplicated all analyses with a decreased sensitivity of $p=0.75$. These data can be found in the supplementaries in~\ref{S1_Fig} and~\ref{S2_Fig}.

In the following section, we provide on overview of the different pooling strategies. Critically, in this article we address the \textit{specific} scenario of a mass screening campaign of an entire population, which during the course of a pandemic likely will have to be repeated several times. In this scenario - and especially in the case of an emerging pathogen -, it can be assumed that neither the number of test kits nor the logistics available would be sufficient for individual testing. These assumptions are also supported by our calculations presented in the Results section. Given the explicit goal of this study, focusing on sensitivity and specificity alone provides an incomplete picture as it ignores an urgent public health priority, which is mass testing of sufficiently large parts of the population. This study therefore employs ``confirmed cases per test'' as main parameter for optimisation. It follows the general idea that if we cannot find all cases in a screening campaign, then we should still aim to find as many cases as possible using all test kits available in a certain time period.

Given the nature of the work, the study was exempted from ethics approval.

\subsection*{Testing strategies}

\begin{figure}[ht]
\begin{subfigure}[t]{.4\textwidth}
    \centering
    \resizebox{\linewidth}{!}{
        \begin{tikzpicture}[
sample/.style={
rectangle,
rounded corners,
very thick,
draw=red!100!black!40,
top color=white,
bottom color=red!100!black!40,
},
mycircle/.style={
circle,
inner sep=0.5mm,
thick,
draw=black,%
bottom color= black,
}
,
mygraph/.style={
grow right=10pt, simple,
}
]
\matrix[row sep = 20pt, column sep = 8pt]{
    \node  [sample, bottom color=blue!60!black!50, draw=blue!60!black!50] (s1) {
        \tikz[
        gray, thick, every node/.style={mycircle}, every graph/.style={mygraph}
        ]
        \graph [grow right=8pt, simple] {
        
        p0/"";
        };
	};
	&
	\node  [sample] (s2) {
        \tikz[
        gray, thick, every node/.style={mycircle}
        ]
        \graph [grow right=8pt, simple] {
        
        p1/""[red, bottom color = red, top color= white];
        };
	};
	&
	\node  [sample, bottom color=blue!60!black!50, draw=blue!60!black!50] (s3) {
        \tikz[
        gray, thick, every node/.style={mycircle}
        ]
        \graph [grow right=8pt, simple] {
        
        p2/"";
        };
	};
	&
	\node  [sample, bottom color=blue!60!black!50, draw=blue!60!black!50] (s4) {
        \tikz[
        gray, thick, every node/.style={mycircle}
        ]
        \graph [grow right=8pt, simple] {
        
        p3/"";
        };
	};
	&
	\node  [sample] (s5) {
        \tikz[
        gray, thick, every node/.style={mycircle}
        ]
        \graph [grow right=8pt, simple] {
        
        p4/"" [red, bottom color = red, top color= white];
        };
	};
	&
	\node  [sample, bottom color=blue!60!black!50, draw=blue!60!black!50] (s6) {
        \tikz[
        gray, thick, every node/.style={mycircle}
        ]
        \graph [grow right=8pt, simple] {
        
        p5/"";
        };
	};
	&
	\node  [sample, bottom color=blue!60!black!50, draw=blue!60!black!50] (s7) {
        \tikz[
        gray, thick, every node/.style={mycircle}
        ]
        \graph [grow right=8pt, simple] {
        
        p6/"";
        };
	};
	&
	\node  [sample, bottom color=blue!60!black!50, draw=blue!60!black!50] (s8) {
        \tikz[
        gray, thick, every node/.style={mycircle}
        ]
        \graph [grow right=8pt, simple] {
        
        p7/"";
        };
	};
	\\
};
\graph[use existing nodes]{

s1;
s2;
s3;
s4;
s5;
s6;
s7;
s8;
};

\end{tikzpicture}
    }
    \caption{individual testing}
    \label{fig:ind}
\end{subfigure}
\begin{subfigure}[t]{.48\textwidth}
    \centering
	\includegraphics[width=0.6\linewidth]{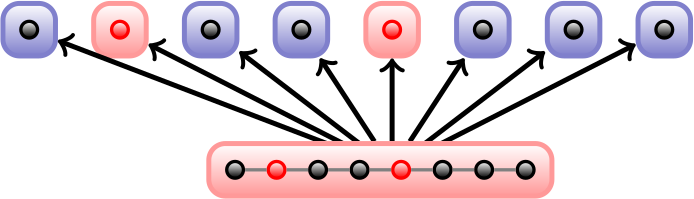}
    \caption{2-level pooling}
    \label{fig:2l}
\end{subfigure}
\begin{subfigure}[t]{.48\textwidth}
    \centering
    \resizebox{\linewidth}{!}{
        \begin{tikzpicture}[
sample/.style={
rectangle,
rounded corners,
very thick,
draw=red!100!black!40,
top color=white,
bottom color=red!100!black!40,
},
mycircle/.style={
circle,
inner sep=0.5mm,
thick,
draw=black,%
bottom color= black,
}
,
mygraph/.style={
grow right=10pt, simple,
}
]
\matrix[row sep = 20pt, column sep = 8pt]{
    &
    \node  [sample, bottom color=blue!60!black!50, draw=blue!60!black!50] (s1) {
        \tikz[
        gray, thick, every node/.style={mycircle}, every graph/.style={mygraph}
        ]
        \graph [grow right=8pt, simple] {
        
        p0/"";
        };
	};
	&
	\node  [sample] (s2) {
        \tikz[
        gray, thick, every node/.style={mycircle}
        ]
        \graph [grow right=8pt, simple] {
        
        p1/""[red, bottom color = red, top color= white];
        };
	};
	& &
	\node  [sample] (s3) {
        \tikz[
        gray, thick, every node/.style={mycircle}
        ]
        \graph [grow right=8pt, simple] {
        
        p5/""[red, bottom color = red, top color= white];
        };
	};
	&
	\node  [sample, bottom color=blue!60!black!50, draw=blue!60!black!50] (s4) {
        \tikz[
        gray, thick, every node/.style={mycircle}
        ]
        \graph [grow right=8pt, simple] {
        
        p6/"";
        };
	};
	&
	\\
	&
	\node  [sample] (s5) {
        \tikz[
        gray, thick, every node/.style={mycircle}
        ]
        \graph [grow right=8pt, simple] {
        
        p0/"" -- p1/""[red, bottom color = red, top color= white];
        };
	};
	&
	\node  [sample, bottom color=blue!60!black!50, draw=blue!60!black!50] (s6) {
        \tikz[
        gray, thick, every node/.style={mycircle}
        ]
        \graph [grow right=8pt, simple] {
        
        p2/"" -- p3/"";
        };
	};
	& &
	\node  [sample] (s7) {
        \tikz[
        gray, thick, every node/.style={mycircle}
        ]
        \graph [grow right=8pt, simple] {
        
        p4/""[red, bottom color = red, top color= white] --  p5/"";
        };
	};
	&
	\node  [sample, bottom color=blue!60!black!50, draw=blue!60!black!50] (s8) {
        \tikz[
        gray, thick, every node/.style={mycircle}
        ]
        \graph [grow right=8pt, simple] {
        
        p6/"" [black, bottom color= black] -- p7/"";
        };
	};
	& &
	\\
	&
	&
	\node  [sample] (s9) {
        \tikz[
        gray, thick, every node/.style={mycircle}
        ]
        \graph [grow right=8pt, simple] {
        p0/"" -- p1/""[red, bottom color = red, top color= white]  -- p2/"" -- p3/"";
        };
	};
	&
	&
	\node  [sample] (s10) {
        \tikz[
        gray, thick, every node/.style={mycircle}
        ]
        \graph [grow right=8pt, simple] {
        
        p4/""[red, bottom color = red, top color= white] --  p5/"" --  p6/"" [black, bottom color= black] -- p7/""];
        };
	};
	&
	&
	\\
	& & &
	\node  [sample] (s11) {
        \tikz[
        gray, thick, every node/.style={mycircle}
        ]
        \graph [grow right=10pt, simple] {
        p0/"" -- p1/""[red, bottom color = red, top color= white]  -- p2/"" -- p3/"" -- p4/""[red, bottom color = red, top color= white] --  p5/"" --  p6/"" [black, bottom color= black] -- p7/"";
        };
	};
	& & 
	\\
};
\graph[use existing nodes]{

s1 <-[very thick] s5;
s2 <-[very thick] s5;

s3 <-[very thick] s7;
s4 <-[very thick] s7;

s5 <-[very thick] s9;
s6 <-[very thick] s9;
s7 <-[very thick] s10;
s8 <-[very thick] s10;

s9 <-[very thick] s11;
s10 <-[very thick] s11;

};

\end{tikzpicture}
    }
    \caption{binary splitting}
    \label{fig:bs}
\end{subfigure}
\begin{subfigure}[t]{.48\textwidth}
    \centering
    \resizebox{\linewidth}{!}{
        \begin{tikzpicture}[
sample/.style={
rectangle,
rounded corners,
very thick,
draw=red!100!black!40,
top color=white,
bottom color=red!100!black!40,
},
mycircle/.style={
circle,
inner sep=0.5mm,
thick,
draw=black,%
bottom color= black,
}
,
mygraph/.style={
grow right=10pt, simple,
}
]
\matrix[row sep = 20pt, column sep = 8pt]{
    &
    \node [sample, bottom color=blue!60!black!50, draw=blue!60!black!50] (s1) {
        \tikz[
        gray, thick, every node/.style={mycircle}, every graph/.style={mygraph}
        ]
        \graph [grow right=8pt, simple] {
        
        p0/"";
        };
	};
	&
	\node  [sample] (s2) {
        \tikz[
        gray, thick, every node/.style={mycircle}
        ]
        \graph [grow right=8pt, simple] {
        
        p1/""[red, bottom color = red, top color= white];
        };
	};
	&
	\node [sample, bottom color=blue!60!black!50, draw=blue!60!black!50] (s3) {
        \tikz[
        gray, thick, every node/.style={mycircle}
        ]
        \graph [grow right=8pt, simple] {
        
        p2/"" -- p3/"";
        };
	};
	&
	\node  [sample] (s3a) {
        \tikz[
        gray, thick, every node/.style={mycircle}
        ]
        \graph [grow right=8pt, simple] {
        
        p4/""[red, bottom color = red, top color= white];
        };
	};
	&
	\\
	& &
	\node  [sample] (s5) {
        \tikz[
        gray, thick, every node/.style={mycircle}
        ]
        \graph [grow right=8pt, simple] {
        
        p0/"" -- p1/""[red, bottom color = red, top color= white];
        };
	};
	& 
		\node [sample, bottom color=white, draw=black] (s6) {
        \tikz[
        gray, thick, every node/.style={mycircle}
        ]
        \graph [grow right=8pt, simple] {
        
        p0/"" -- p1/""[black, bottom color = black];
        };
	};
	&
	\node  [sample] (s7) {
        \tikz[
        gray, thick, every node/.style={mycircle}
        ]
        \graph [grow right=8pt, simple] {
        
        p2/"" -- p3/"" -- p4/""[red, bottom color = red, top color= white];
        };
	};
	&
	\node  [sample, bottom color=blue!60!black!50, draw=blue!60!black!50] (s8) {
        \tikz[
        gray, thick, every node/.style={mycircle}
        ]
        \graph [grow right=8pt, simple] {
        
        p5/"" -- p6/"" [black, bottom color= black] -- p7/"";
        };
	};
	& &
	\\
	& &
	&
	\node  [sample] (s9) {
        \tikz[
        gray, thick, every node/.style={mycircle}
        ]
        \graph [grow right=8pt, simple] {
        p0/"" -- p1/""[red, bottom color = red, top color= white]  -- p2/"" -- p3/"";
        };
	};
	&
	\node  [sample] (s10) {
        \tikz[
        gray, thick, every node/.style={mycircle}
        ]
        \graph [grow right=8pt, simple] {
        
        p2/"" -- p3/"" -- p4/""[red, bottom color = red, top color= white] --  p5/"" --  p6/"" [black, bottom color= black] -- p7/""];
        };
	};
	&
	\node [sample, bottom color=white, draw=black] (s10b) {
        \tikz[
        gray, thick, every node/.style={mycircle}, every graph/.style={mygraph}
        ]
        \graph [grow right=8pt, simple] {
        
        p0/""[red, bottom color = red, top color= white] -- p1/"" -- p2/"" -- p3/"";
        };
	};
	&
	\\
	& & & 
	&
	\node  [sample] (s11) {
        \tikz[
        gray, thick, every node/.style={mycircle}
        ]
        \graph [grow right=10pt, simple] {
        p0/"" -- p1/""[red, bottom color = red, top color= white]  -- p2/"" -- p3/"" -- p4/""[red, bottom color = red, top color= white] --  p5/"" --  p6/"" [black, bottom color= black] -- p7/"";
        };
	};
	& & 
	\\
};
\graph[use existing nodes]{

s3 <-[very thick] s7;
s3a <-[thick, dashed] s3;
s3a <-[thick, dashed] s7;

s7 <-[very thick] s10;
s8 <-[very thick] s10;

s6 <-[thick, dotted] s1;
s6 <-[thick, dotted] s2;
s6 <-[thick, dotted] s9;
s10 <-[thick, dotted] s6;

s10b <-[thick, dotted] s11;
s10 <-[thick, dotted] s10b;


s2 <-[thick, dashed] s1;
s2 <-[thick, dashed] s5;
s1 <-[very thick] s5;

s5 <-[very thick] s9;

s9 <-[very thick] s11;

s10b;
};

\end{tikzpicture}
    }
    \caption{recursive binary splitting}
    \label{fig:rbs}
\end{subfigure}
\begin{subfigure}[t]{.48\textwidth}
    \centering
    \includegraphics[width=0.6\linewidth]{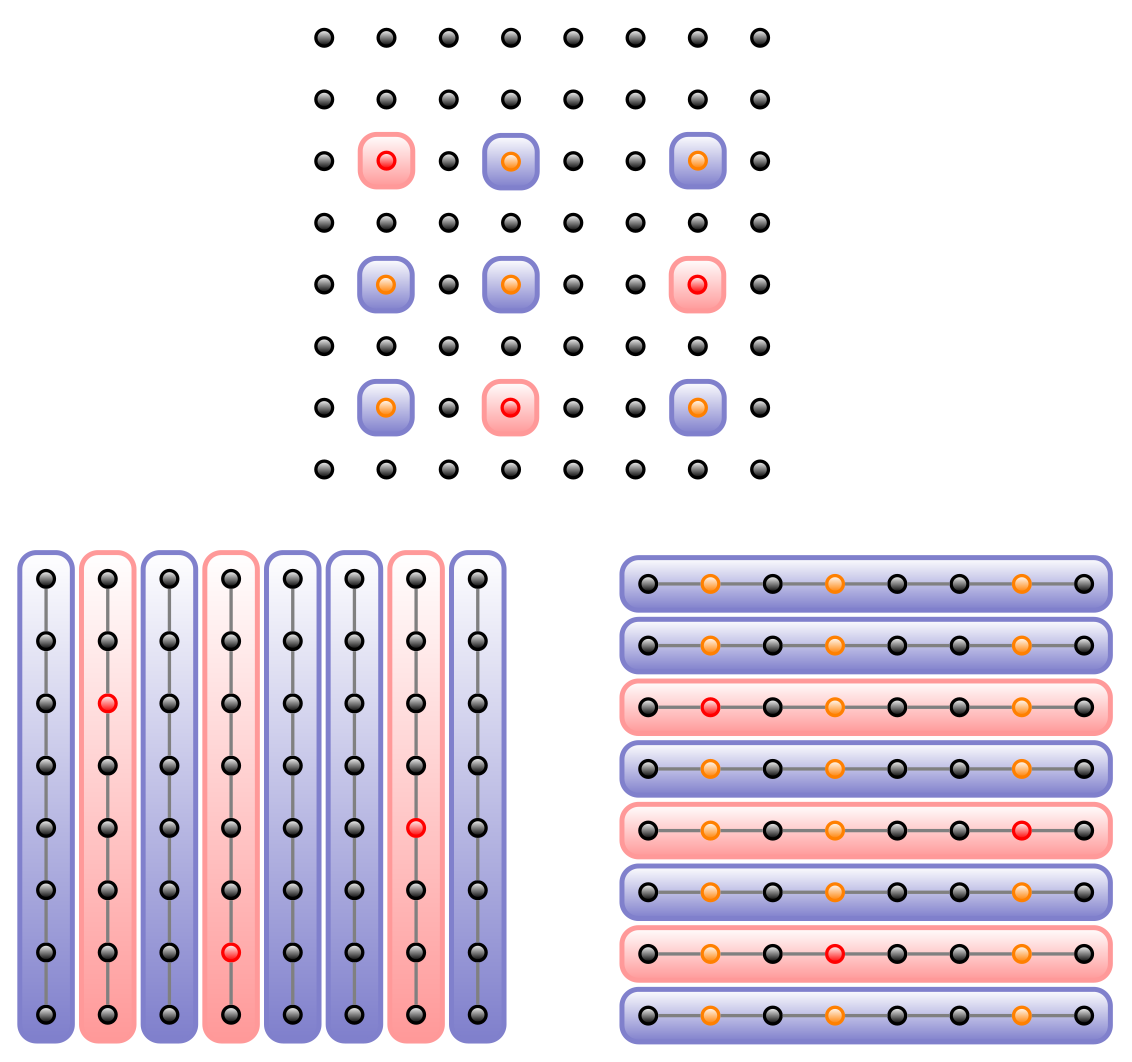}
    \caption{Purim}
    \label{fig:purim}
\end{subfigure}
\begin{subfigure}[t]{.48\textwidth}
    \centering
    \resizebox{\linewidth}{!}{
        \begin{tikzpicture}[
sample/.style={
rectangle,
rounded corners,
very thick,
draw=red!100!black!40,
top color=white,
bottom color=red!100!black!40,
},
mycircle/.style={
circle,
inner sep=0.5mm,
thick,
draw=black,%
bottom color= black,
}
,
mygraph/.style={
grow right=10pt, simple,
}
]
\matrix[row sep = 20pt, column sep = 8pt]{
    & & & 
    \node  [sample] (sa) {
        \tikz[
        gray, thick, every node/.style={mycircle}
        ]
        \graph [grow right=8pt, simple] {
        
        p1/""[red, bottom color = red, top color= white];
        };
	};
	& 
    \node [sample, bottom color=white, draw=black] (sb) {
        \tikz[
        gray, thick, every node/.style={mycircle}, every graph/.style={mygraph}
        ]
        \graph [grow right=8pt, simple] {
        
        p0/"";
        };
	};
	&
	\node [sample, bottom color=blue!60!black!50, draw=blue!60!black!50] (s4b) {
        \tikz[
        gray, thick, every node/.style={mycircle}, every graph/.style={mygraph}
        ]
        \graph [grow right=8pt, simple] {
        
        p0/"" -- p1/"" -- p2/"" -- p3/"" --p4/"";
        };
	};
	
	\\
    &
    \node [sample, bottom color=blue!60!black!50, draw=blue!60!black!50] (s1) {
        \tikz[
        gray, thick, every node/.style={mycircle}, every graph/.style={mygraph}
        ]
        \graph [grow right=8pt, simple] {
        
        p0/"";
        };
	};
	&
	\node  [sample] (s2) {
        \tikz[
        gray, thick, every node/.style={mycircle}
        ]
        \graph [grow right=8pt, simple] {
        
        p1/""[red, bottom color = red, top color= white];
        };
	};
	& &
	\node  [sample] (s3) {
        \tikz[
        gray, thick, every node/.style={mycircle}
        ]
        \graph [grow right=8pt, simple] {
        
        p4/""[red, bottom color = red, top color= white] -- p5/"";
        };
	};
	&
	\node [sample, bottom color=white, draw=black] (s4) {
        \tikz[
        gray, thick, every node/.style={mycircle}, every graph/.style={mygraph}
        ]
        \graph [grow right=8pt, simple] {
        
        p0/"" -- p1/"" -- p2/"" -- p3/"";
        };
	};
	&
	\\
	& &
	\node  [sample] (s5) {
        \tikz[
        gray, thick, every node/.style={mycircle}
        ]
        \graph [grow right=8pt, simple] {
        
        p0/"" -- p1/""[red, bottom color = red, top color= white];
        };
	};
	& 
	\node [sample, bottom color=white, draw=black] (s6) {
        \tikz[
        gray, thick, every node/.style={mycircle}
        ]
        \graph [grow right=8pt, simple] {
        
        p0/"" -- p1/""[black, bottom color = black];
        };
	};
	& 
	\node  [sample] (s7) {
        \tikz[
        gray, thick, every node/.style={mycircle}
        ]
        \graph [grow right=8pt, simple] {
        
        p2/""[red, bottom color = red, top color= white] -- p3/"" -- p4/"" [black, bottom color= black]--  p5/"" -- p6/"" -- p7/"" ];
        };
	};
& &
	\\
	& &
	&
	\node  [sample] (s9) {
        \tikz[
        gray, thick, every node/.style={mycircle}
        ]
        \graph [grow right=8pt, simple] {
        p0/"" -- p1/""[red, bottom color = red, top color= white]  -- p2/"" -- p3/"";
        };
	};
	&
	\node [sample, bottom color=white, draw=black] (s10) {
        \tikz[
        gray, thick, every node/.style={mycircle}
        ]
        \graph [grow right=8pt, simple] {
        
        p2/""[red, bottom color = red, top color= white] -- p3/"" -- p4/"" [black, bottom color= black]--  p5/"" ];
        };
	};
	&
	&
	\\
	& & & &
	\node  [sample] (s11) {
        \tikz[
        gray, thick, every node/.style={mycircle}
        ]
        \graph [grow right=10pt, simple] {
        p0/"" -- p1/""[red, bottom color = red, top color= white]  -- p2/"" -- p3/"" -- p4/""[red, bottom color = red, top color= white] --  p5/"" --  p6/"" [black, bottom color= black] -- p7/"";
        };
	};
	& & 
	\\
};
\graph[use existing nodes]{

s3 <-[very thick] s7;
s10 <-[thick, dotted] s11;

s2 <-[thick, dashed] s1;
s2 <-[thick, dashed] s5;
s1 <-[very thick] s5;

s5 <-[very thick] s9;
s6 <-[thick, dotted] s9;

s7 <-[thick, dotted] s6;
s7 <-[thick, dotted] s10;
s6 <-[thick, dotted] s2;
s6 <-[thick, dotted] s1;

s4 <-[thick, dotted] s7;
s4b <-[thick, dotted] s4;

s9 <-[very thick] s11;

sa <-[very thick] s3;
sb <-[thick, dotted] sa;
sb <-[thick, dotted] s3;

s4b <-[thick, dotted] sb;

};

\end{tikzpicture}
    }
    \caption{Sobel-R1}
    \label{fig:sobel}
\end{subfigure}
\caption{{\bf An illustration of finding cases in a pool} Demonstrated for a pool of eight samples for individual testing (A), 2-level pooling (B), binary splitting (C), and Sobel-R1 (F). For recursive binary splitting (D), we show part of the search tree highlighting the difference to binary splitting. For Purim (E), we show the search for three cases in an overall set of $8^2=64$, starting from 16 pools of size eight (i.e.\ 8 horizontal and 8 vertical in a matrix arrangement) before testing the cross-sections.
}
\label{fig1}
\end{figure}
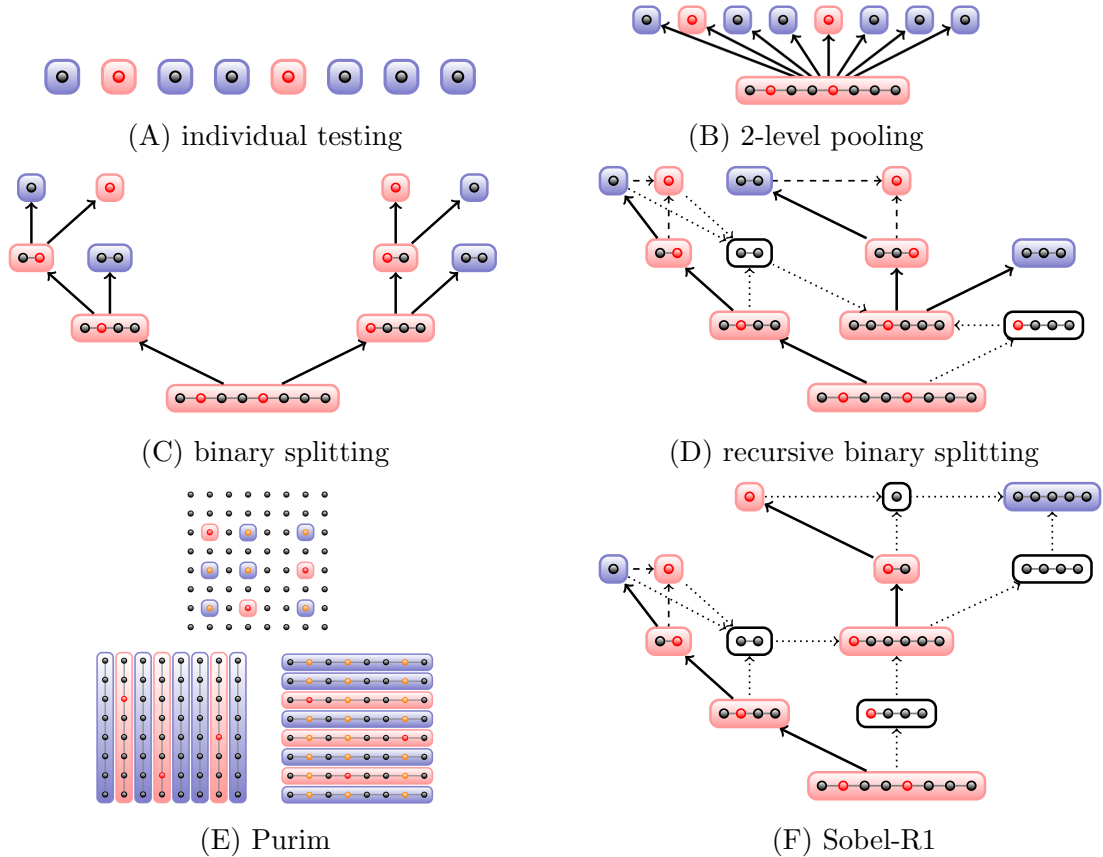

We summarise the methods and illustrate them in Fig~\ref{fig1}.

\begin{description}
	\item[Individual testing] The conventional approach of testing every person in a given population individually.
	
	\item[2-level pooling] Following a recent preprint by Hanel and Thurner we define a maximum pool size $k$. If the pooled test is positive, every sample in that pool is tested individually.\cite{AustrianPaper} Specifically, if the first level test of the entire pool is negative, then the whole pool is cleared, without need for individual testing.
	This procedure was first introduced by Dorfman in 1943 and improved by Sterret in 1957.\cite{Dorfman, Sterrett}
	
	\item[Binary splitting] A well-known hierarchical multi-layer procedure: if a test of a pool size $k$ is positive, the pool is split in two sets of size $k/2$, and a pooled test is performed on the two new sets.\cite{MiltonGroll:GroupTesting}
	If a pool is tested negative, it is cleared and no further tests are performed on the samples in this pool. 
	This procedure is repeated recursively for those subsets with a positive (pooled) test until each individual case has been identified.
	
	\item[Optimised recursive binary splitting] A recent variation of binary splitting: if at a given level of the hierarchy only one pool tests positive, then the identification of a particular case continues via a binary search. Afterwards, all confirmed positive or confirmed negative samples are removed from the pool and the procedure continues with the remaining subjects in the unified pool.\cite{RBS}
	We improved the method by choosing optimal initial pool sizes based on the infection rate.
	
	Each level of the optimised recursive binary splitting (oRBS) procedure is executed as follows: 
	\begin{enumerate}
	    \item  Starting with a pool $P$ of (optimised) size $k$ and testing the entire pool, unless it is know from a previous step of the algorithm that $P$ must contain a positive sample.
	    \item  If $P$ is tested positive, then it is split into partitions $P_1$ and $P_2$ and tests are performed on these two new pools.
	    \item If both $P_1$ and $P_2$ are positive, the oRBS algorithm is run recursively on $P_1$ and $P_2$.
	If only one partition, say $P_1$, tests positive, then all individuals in $P_2$ are identified as healthy and a binary splitting subprocedure is run on $P_1$.
	
		The purpose of this refinement step is to identify a single case in the pool $P_1$ and to remove this case along with any individuals that have been identified as healthy.
		\item Finally, we run the oRBS algorithm recursively on all remaining samples in $P_1$.
	\end{enumerate}
	
	Consider the example step in Fig~\ref{fig1}(D). The starting point is a positive sample pool $P_1$ of size $k = 8$ from a previous oRBS step. Executing the binary subprocedure to identify a first case in $P_1$, the pool is split into two subpools $S_1$ and $S_2$ of size $4$. To identify this case, it is sufficient to perform a test on one of the subpools. In our example, the tested subpool $S_1$ contains one case, leading to another binary split. Only one of the resulting pools $S_{1,1}$ and $S_{1,2}$ of size $2$ is tested. $S_{1,1}$ is identified as positive and split into two single samples. One of these two samples is tested negative, implying that the second sample has to be positive.
	
	At this point we have identified one case and one healthy individual in the pool $P_1$. Removing these two individuals, the oRBS algorithm is performed analogously on the remaining six samples in $P_1$ until all cases in $P_1$ have been identified. 
	
	\item[Purim] A matrix-based pooling approach where one-dimensional overlapping pools are arranged in a matrix, and only cross-sections of pools that have tested positive 
	are tested individually.\cite{PURIM}
    
    In the example illustrated in Fig~\ref{fig1}(E), we consider a set of 64 individuals to be tested and choose a fixed pool size of $k=8$. The samples are arranged into a $8 \times 8$ matrix and the eight pools formed by the columns of this matrix are tested. As a next step, the eight pools formed by the rows of the matrix are tested, followed by individual tests on samples where both column and row have tested positive.
    
    We only consider Purim's 2D variant and neglect the 3D variant; the latter becomes impractical for low infection rates, requiring handling of up to $32^3=32,768$ samples simultaneously.
	
	\item[Sobel-R1] A decision tree approach based on the assumption of a binomial distribution of the test results. Pool sizes are adapted according to the minimisation of the expected number of remaining tests.\cite{MiltonGroll:GroupTesting}
	
	This approach assumes that all $n$ individuals to be tested are partitioned into (at most) two sets; a set of size $m \ge 0$ containing at least one person that is known to be infected, and a binomial set of size $n-m \ge 0$. 
	Denote by $q$ the probability of being tested negative and by $k$ the next pool size we need to determine.
	The expected number of pooled tests remaining to be performed for a set of size $m$ and a binomial set of size $n-m$ is given recursively by
	\begin{align*}
	    G(0, n) \ = \ 1 + \min_{1 \le k \le n} \left\lbrace q^k G(0, n-k) + (1-q^k) G(k, n)\right\rbrace,
	\end{align*}
	for $m = 0$, and
	\begin{align*}
	    G(m, n) \ = \ 1 + \min_{1 \le k \le n} \left\lbrace \left( \frac{q^k - q^m}{1 - q^m} \right) G(m-k,n-k) + \left( \frac{1-q^k}{1-q^m} \right) G(k, n)\right\rbrace,
	\end{align*}
	for $n \ge m \ge 2$, with boundary conditions
	\begin{align*}
	    G(0, 0) \ &= \ 0, \\
	    G(1, n) \ &= \ G(0, n-1) \qquad \text{ for } \qquad n = 1, 2, \ldots \; .
	\end{align*}
	Thus, we can determine the size of the next pool by computing the expected number of remaining tests.
	
	If the infection rate is known, it can be shown that this approach is a stochastically optimal search variant and therefore serves as an upper bound for the number of cases correctly identified per test.
	
	In the example depicted in Fig~\ref{fig1}(F), we start with a pool size $8$ with an infection rate $p = 0.25 = 1-q$. Computing $G(0,8)$ yields $k = 8$, and as a first step the entire pool is tested. Since the test is positive, we continue by computing $G(8,8)$ and inferring that we need to choose $k = 4$ as the size of the next pool to be tested.
	This procedure is repeated until the end of the branch, i.e.\ until a single sample needs to be tested.
	All samples that have not yet been classified are now collected in a new pool, which in our example has size $6$.
	Since computing $G(0,6)$ yields $k = 6$, the entire pool is tested. The result is positive, so we compute $G(6,6)$ and continue analogously to the procedure as described for the previous branch.
	
	The algorithm ends once all samples in the initial pool have been classified as positive or have been found to be negative.

\end{description}

\subsection*{Implementation and statistics}
Samples within the virtual populations are designated as positive randomly using Python's numpy random number generator in accordance with the specified infection rate.
To account for randomness and to obtain robust results, we simulate in repetitions of ten and then calculate the expectation value and standard deviation.

For all pooling approaches we limit the maximum pool size to $32$ which was shown to be a viable upper limit enabling pool-based testing without a significant loss of sensitivity.\cite{IsraelPaper,lohse20pooling} However, for increasing infection rates, smaller pool sizes result in more cases identified per test, which means the initial pool size can be optimised, see e.g.\ Hanel and Thurner and Xiong et al.\cite{AustrianPaper, xiong2019determination}
We optimise the pool size for each testing strategy and infection rate using a test population of $50,000$ individuals using every pool size in $\{1,\ldots,32\}$.
The optimal pool size is the one leading to the fastest screening of the test population.
It is calculated once, stored, and then reused for the appropriate combination of testing method and infection rate.

\begin {figure}[!ht]
\vspace{2em}
    \resizebox{\linewidth}{!}{
    \includegraphics[width=.02\linewidth]{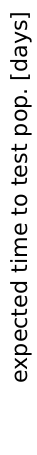}
    \includegraphics[width=.3\linewidth]{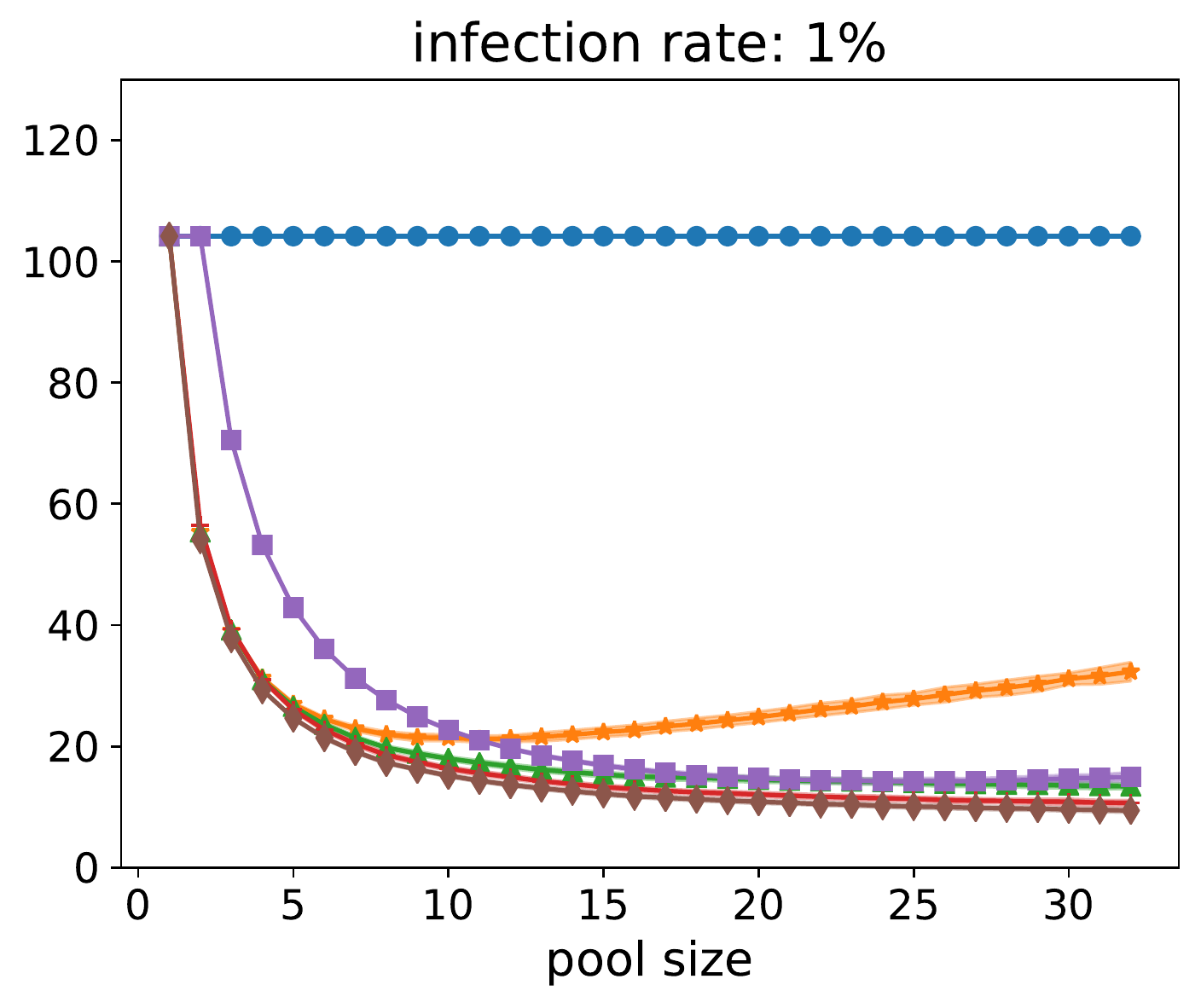}
    \includegraphics[width=.3\linewidth]{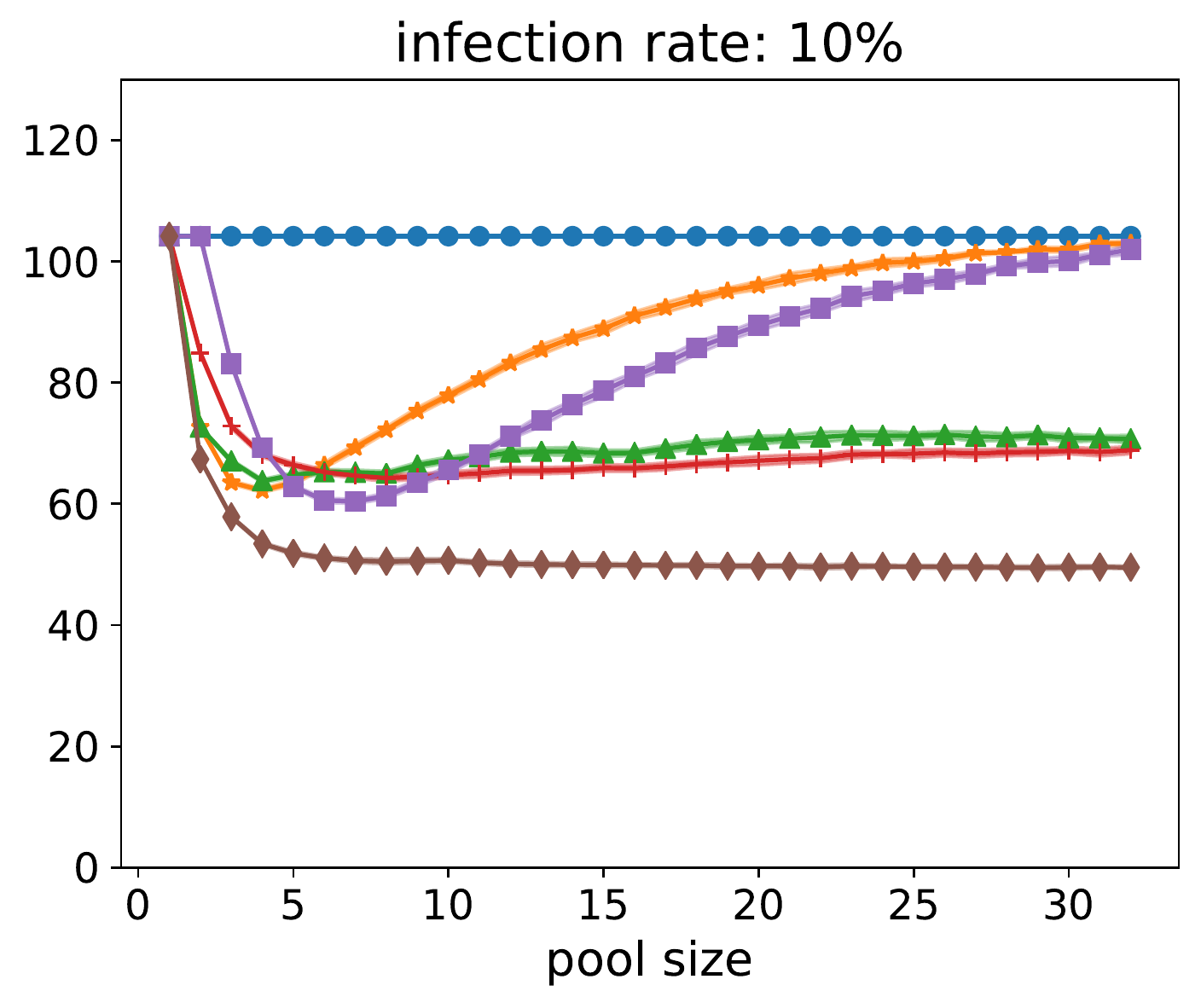}
    \includegraphics[width=.3\linewidth]{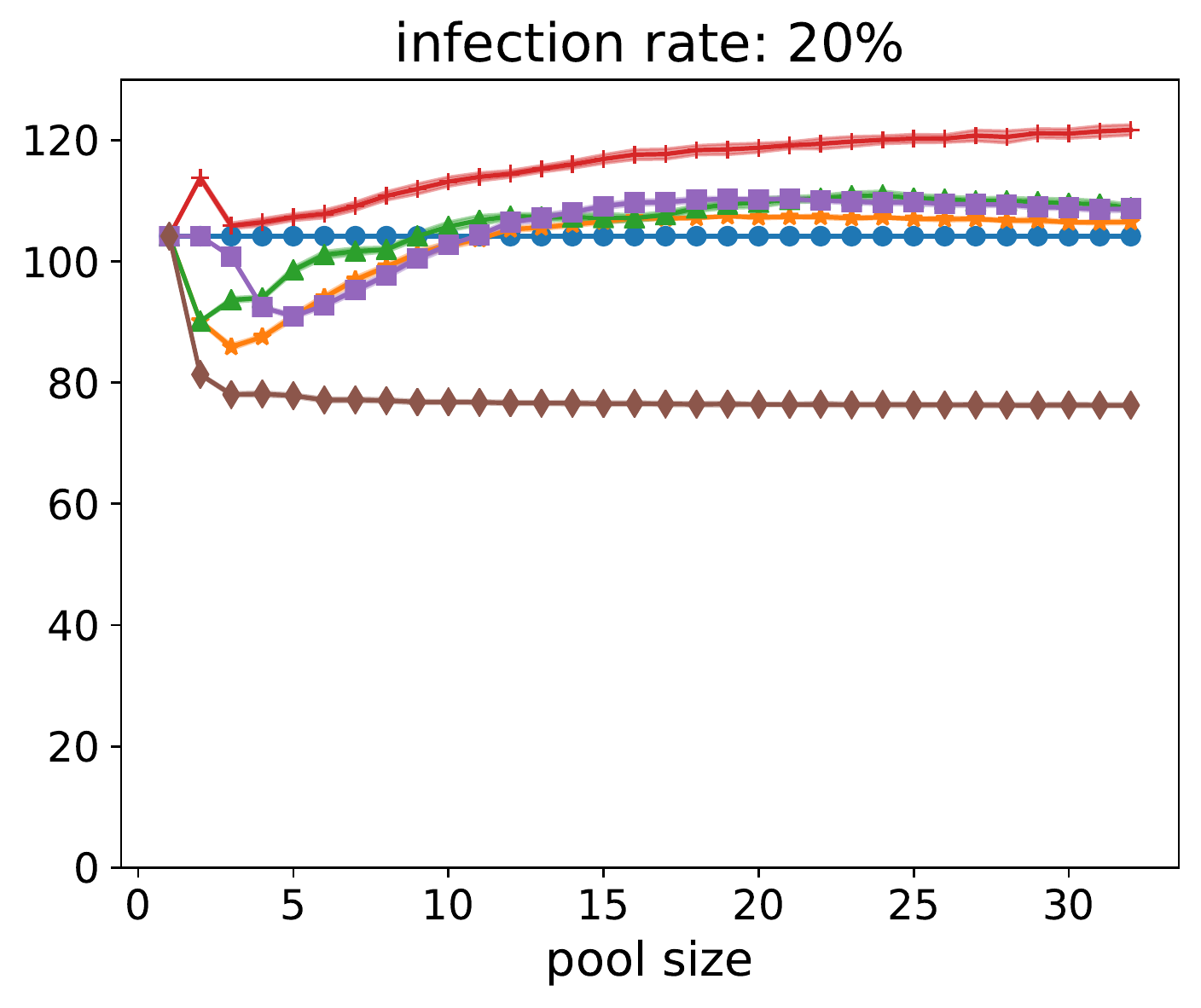}
    }
    \caption{{\bf The best pool size (lowest total time) depends on the infection rate, here for $ir$=1\%, 10\%, 20\%.} For low infection rates, all methods but 2-level pooling prefer pool sizes that are as large as possible. With increasing infection rate, the optimal pool size decreases -- with the exception of the Sobel-R1 method -- until they approach pool size 1. Parameters: sensitivity $p=0.99$, false positive rate $q=0.01$, population $50,000$, test duration 5h, averaged over 10 runs. Blue: individual testing; orange: 2-level pooling; green: binary splitting; red: recursive binary splitting; purple: Purim; brown: Sobel-R1}
    \label{fig2}
\end{figure}

As shown in Fig~\ref{fig2}, the optimal pool size for each method in terms of lowest expected time to test an entire population depends on the infection rate. For an infection rate $ir$=1\%, the expected total time decreases with increasing pool size for all pooling methods except 2-level pooling.
For increasing infection rates, the optimal pool size decreases for all pooling methods but Sobel-R1, and eventually approaches pool size 1 and thus individual testing.

The code and all data are available at
\begin{quote}
    \url{https://github.com/SC-SGS/covid19-pooling}.
\end{quote}

\section{Results}

\label{section:results}

This study employs six different methods to model the most effective strategy to screen a given population within the shortest time and with the smallest number of tests possible. 
Since the choice of ideal testing strategy depends on the input parameters (such as size and infection rate of a population, test sensitivity, etc.), the models we developed can be fully customised and run via 
\begin{quote}
    \url{https://covid19.enfunction.com},  
\end{quote}
enabling the reader to find the best possible strategy for their specific needs.

Fig~\ref{fig3}(A) compares the effectiveness in terms of ICPT for all methods for different infection rates under the assumptions given in the Methods section. Sobel-R1 is the optimal method and provides a theoretical upper bound for the achievable ICPT. However, this method is restricted in its practical applicability (see below). For an infection rate of 1\%, the hierarchical approaches such as binary and recursive binary splitting show almost an eight and ten fold ICPT increase compared to the current status quo of individual testing respectively, enabling the testing of large groups with a single test. Disregarding Sobel-R1, we obtain the following results: for infection rates up to 6\%, recursive binary splitting is the optimal method. For infection rates up to $2.5$\%, the next best option is binary splitting, and for infection rates between $2.5$\% and 6\%, the next best method is the matrix-based Purim method. For infection rates between 6\% and 12\%, Purim is the optimal method. For infection rates of 12\% and higher, 2-level pooling with optimised pool sizes yields the highest ICPT. Of note, the current standard approach of individual testing is never found to be the best choice in the scenarios we modelled.

\begin{figure}[b]
\centering
\begin{subfigure}[t]{\linewidth}
\centering
    \includegraphics[width=0.45\linewidth]{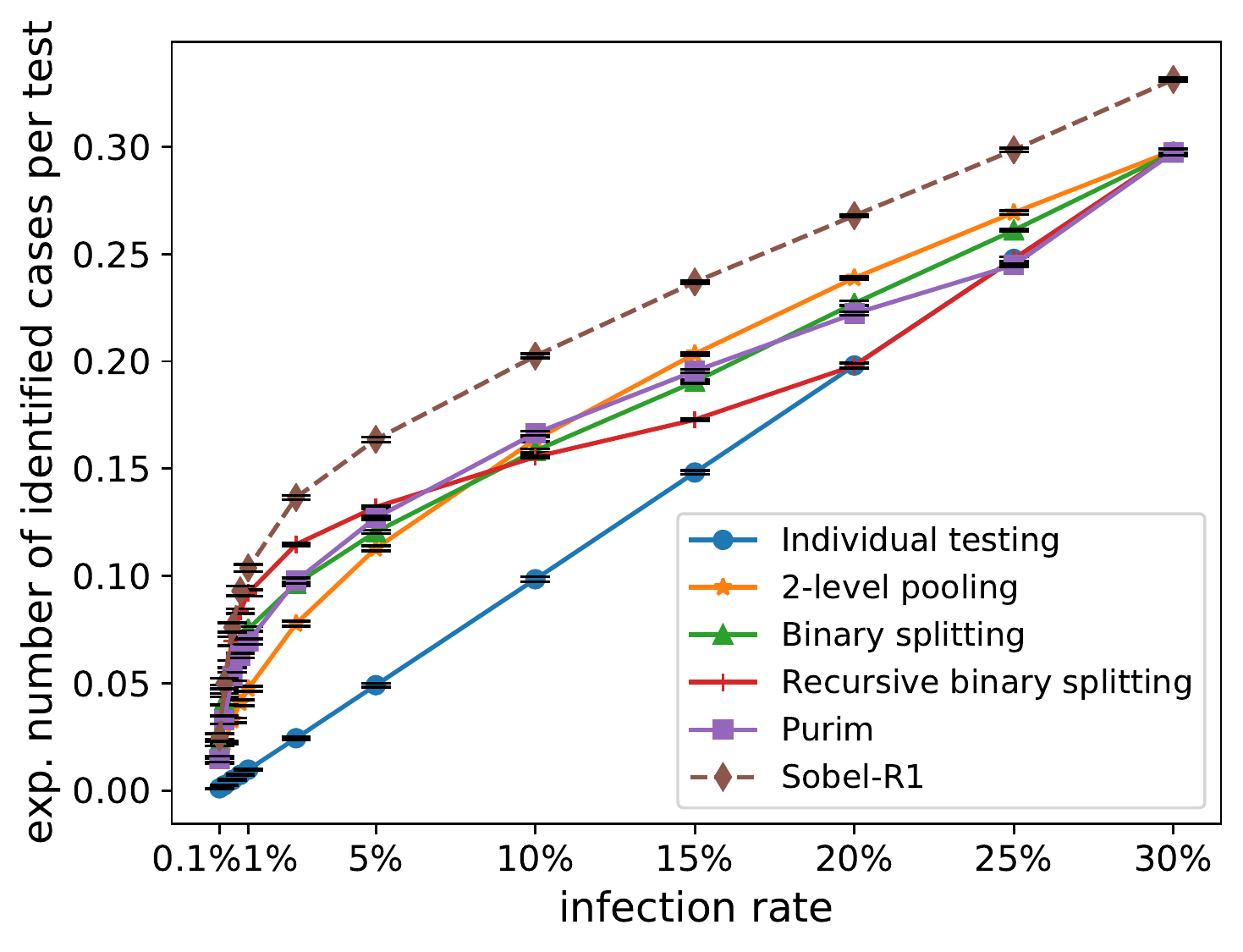}
    \hspace{0.02\linewidth}
    \includegraphics[width=0.45\linewidth]{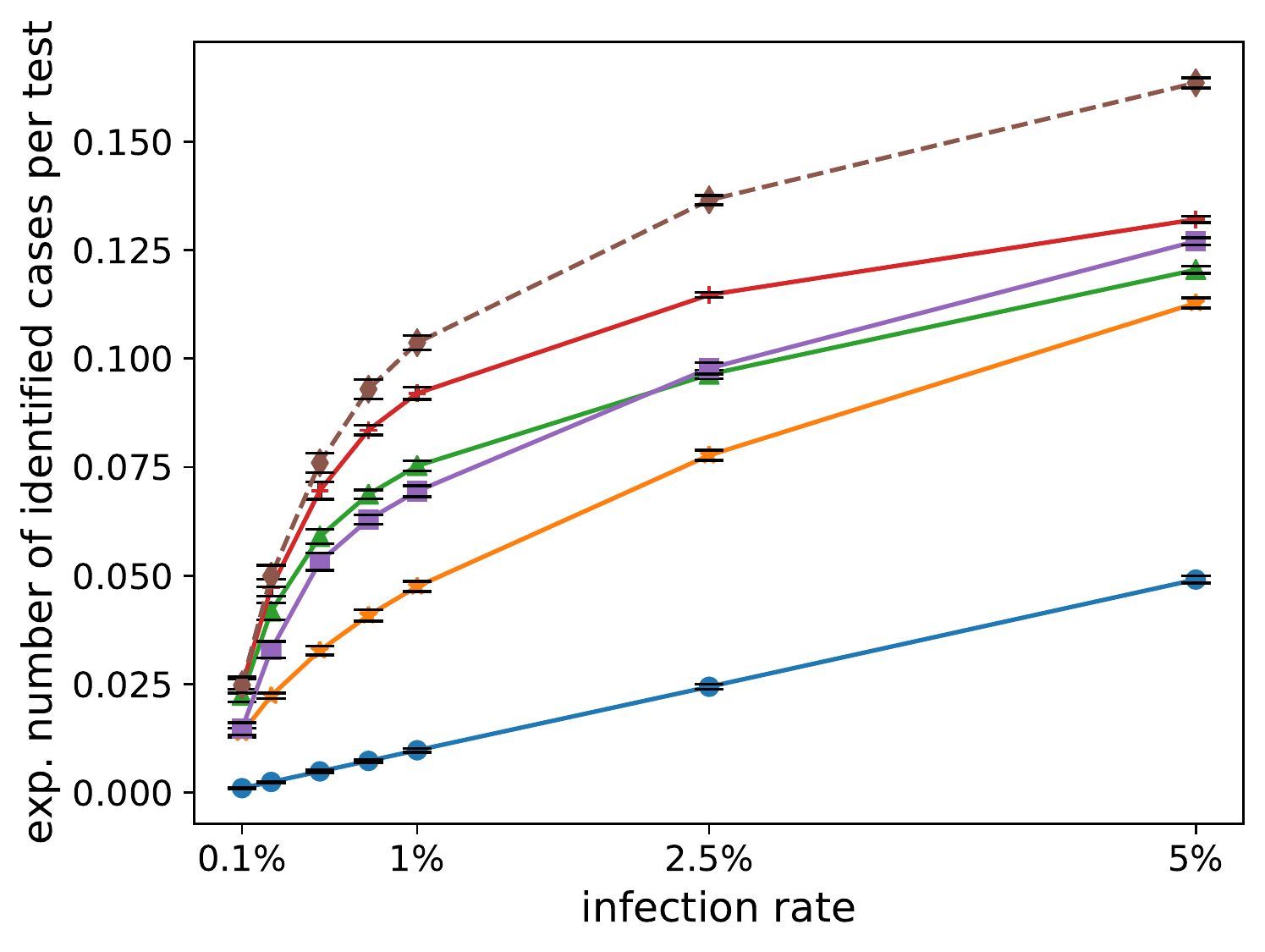}
\caption{Expected number and standard deviation of identified cases per test for different infection rates with optimised group sizes; right: zoom-in for small infection rates} 
    \label{fig:icpt}
\end{subfigure}
\begin{subfigure}[t]{\linewidth}
\centering
    \includegraphics[width=0.45\linewidth]{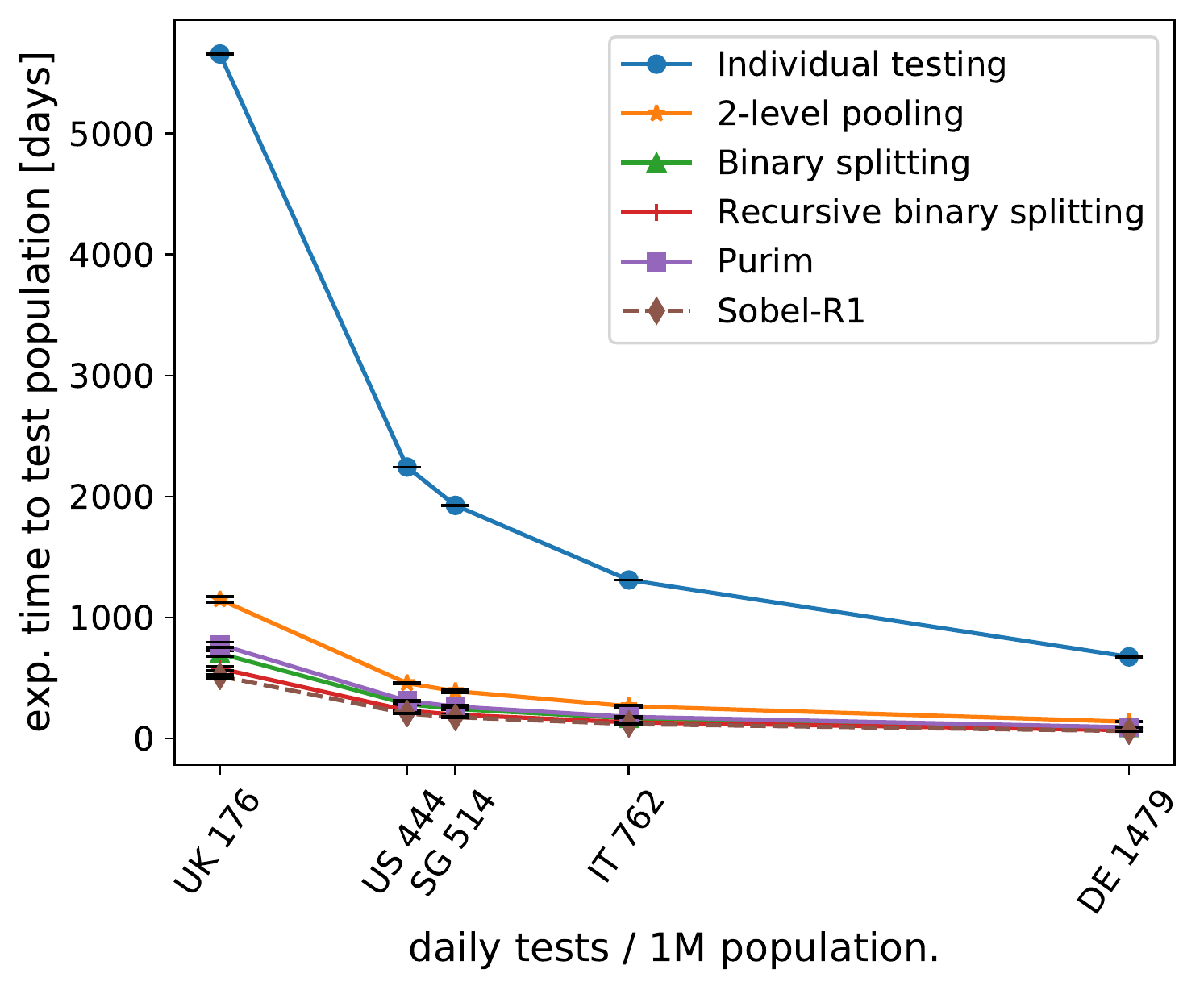}
        \hspace{0.02\linewidth}
    \includegraphics[width=0.45\linewidth]{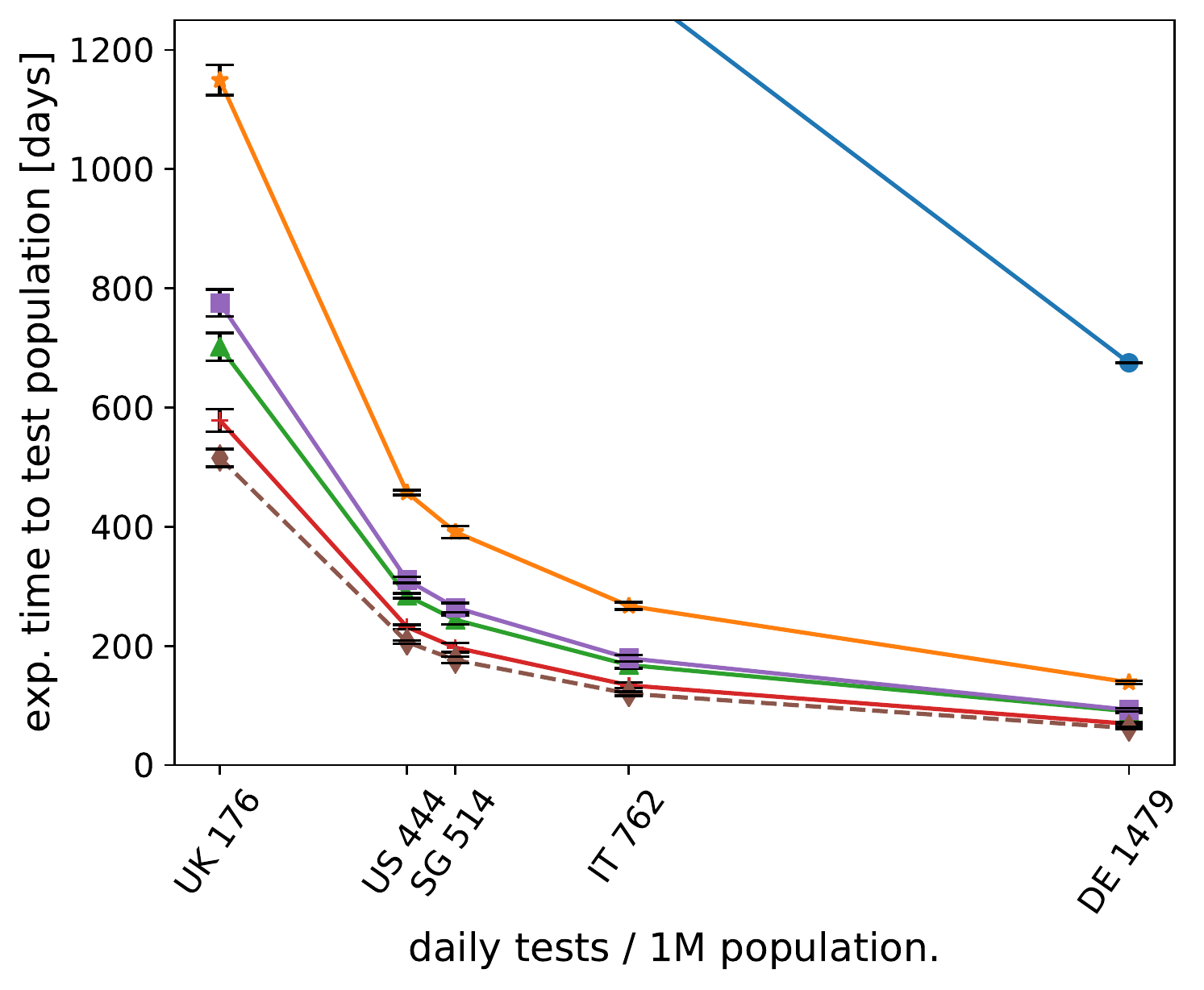}
    \caption{Expected time in days and standard deviation to test the whole population depending on daily test capacity per 1m population, infection rate 1\%; right: zoom-in excl.\ individual testing}
    \label{fig:time_test}
\end{subfigure}
\end{figure}%
\begin{figure}[ht]\ContinuedFloat
\begin{subfigure}[t]{.45\linewidth}
    \includegraphics[width=\linewidth]{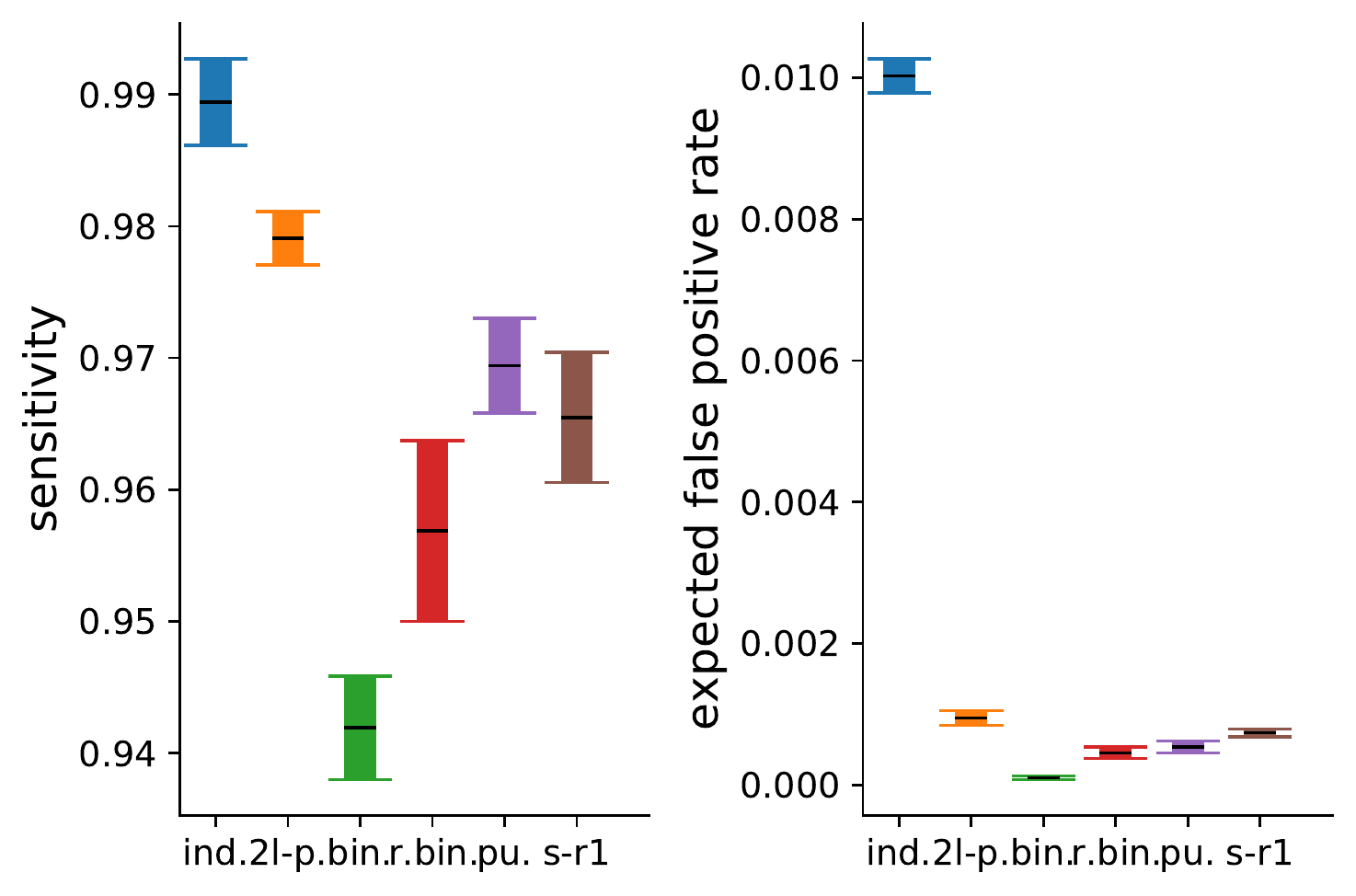}
    \caption{Expected true positives and false positives for the six methods (pop.~100,000)}
    \label{fig:tp_fn} 
\end{subfigure}
\hspace{0.02\linewidth}
\begin{subfigure}[t]{.45\linewidth}
    \includegraphics[width=\linewidth]{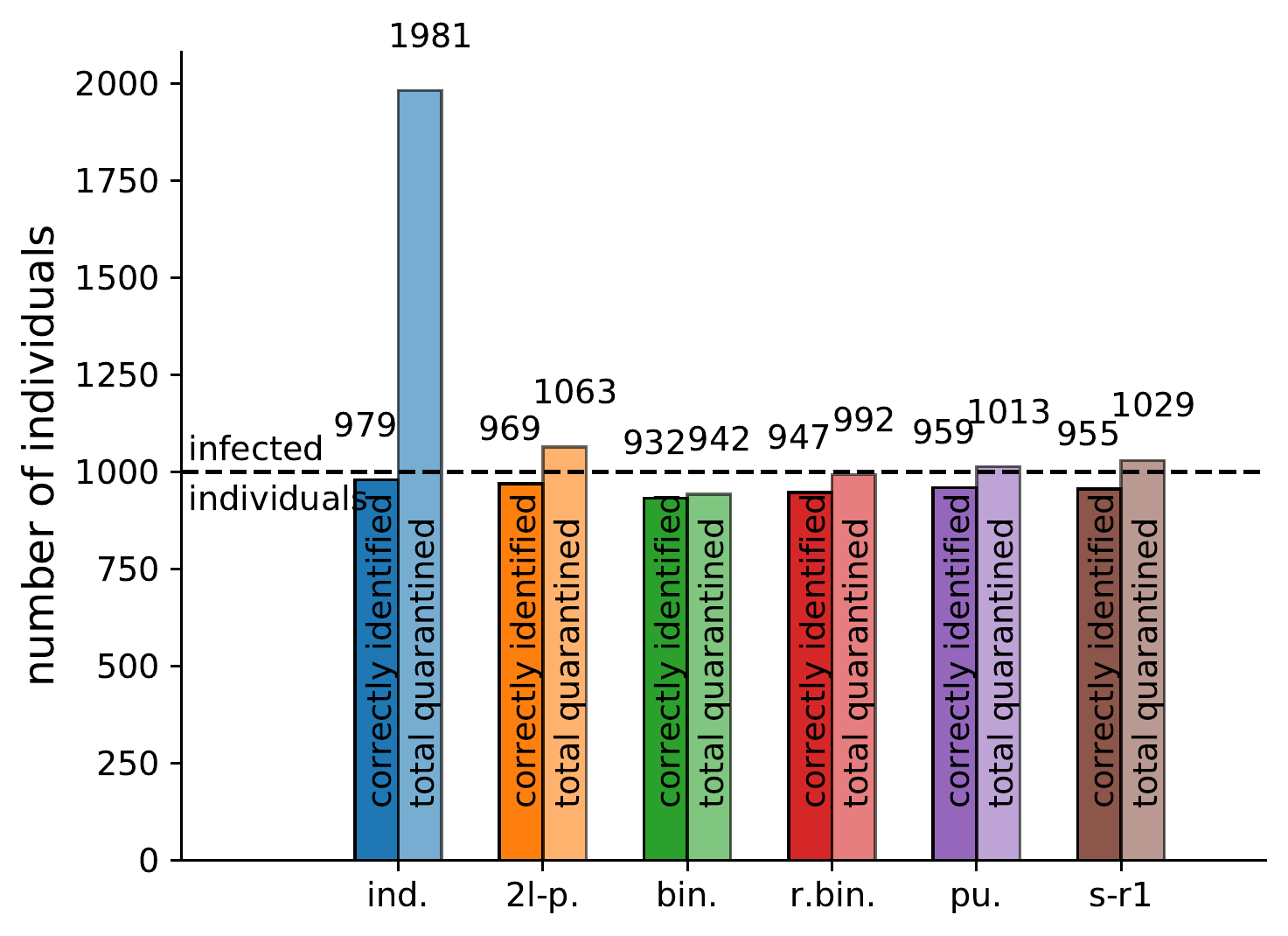} 
    \caption{Numbers of identified cases and quarantined individuals (pop.~100,000)}
    \label{fig:quarantined}
\end{subfigure}
    \label{fig:scenario1}
    \caption{{\bf Screening the whole population.}
Parameters: sensitivity $p=0.99$, false positive rate $q=0.01$, test duration 5h, averaged over 10 runs. Optimal (max.) pool size each (c.f. Fig~\ref{fig2}); for $ir$=1\% as in (B)--(D) we obtain individual testing: 1; 2-level pooling: 12; binary splitting: 32; recursive binary splitting: 32; Purim: 32, Sobel-R1: 31.
}
    \label{fig3}
\end{figure}

To better examine the potential of hierarchical approaches, we deployed our models to simulate the situation for five countries: the US, the UK, Germany, Italy, and Singapore; see Table~\ref{tab:countries}. We assume that as of May 2020 $ir=1\%$ of the population is infected. Fig~\ref{fig3}(B) shows the overall time required to test the whole population in each of the five countries. Even in Germany, which has the highest relative testing capacity per day of these five countries, it would take 675 days with the current screening approach (individual testing) to test every individual; the US on the other hand would require 2,244 days. For the US, binary splitting reduces the time required to 285 days (about 9.5 months), and the optimised recursive binary splitting to 232 days. If only one tenth of the population needs to undergo screening -- e.g.\ when prioritising frontline medical staff and public workers -- then this would mean such a campaign could be completed with binary splitting in about 29 days in the US, 71 in the UK, 25 in Singapore, 17 in Italy, and 10 in Germany. When the infection rates are considerably lower, or when employing an optimal, yet logistically more challenging, pooling method, the gains are more pronounced. With Sobel-R1, the same screening campaign could be completed in 21 days in the US, 52 in the UK, 18 in Singapore, 12 in Italy, and 7 in Germany. As a case study, we conducted a comparison using the six approaches for three different infection rates for the US, see Table~\ref{tab:test10}.

\begin{table}[!ht]
    \centering
    \begin{tabular}{|c+r|r|r|r|}
        \hline
        \textbf{Country} & \textbf{Population} $n$ & \textbf{Daily testing capacity} $c$ & \textbf{Tests per 1m population} \\ \thickhline
        US & 328.24\,m & 146,000 & ~~444\\ \hline
        DE & ~83.15\,m & 123,000 & 1,479 \\ \hline
        UK & ~67.89\,m & ~12,000 & ~~176\\ \hline 
        IT & ~60.31\,m & \,~46,000 & ~~762\\ \hline
        SG & ~~\,5.64\,m & ~~~2,900 & ~~514 \\ \hline
    \end{tabular}
    \caption{\textbf{Data used in the simulations based on country.}
        14 April 2020, various sources (top to bottom: The Atlantic; Robert-Koch-Institute; gov.uk; ourworldindata.org; The Straits Times)
    }
    \label{tab:countries}
\end{table}

\begin{table}[!ht]
    \centering
\begin{tabular}{|l+l|r|r|r|r|r|r|}
    \hline
        \textbf{Method} & \multicolumn{2}{l|}{\bf $ir$=0.1\%} & \multicolumn{2}{l|}{\bf $ir$=1\%} & \multicolumn{2}{l|}{\bf $ir$=2.5\%}\\ \hline
        & \textbf{days} & \textbf{speedup} & \textbf{days} & \textbf{speedup} & \textbf{days} & \textbf{speedup} \\ \thickhline
        Individual testing         & 224.4 &               & 224.4 &               & 224.4 & \\ \hline
        2-level pooling            &  16.4 & 13.7x &  45.8 &  4.9x &  70.1 & 3.2x \\ \hline
        Binary splitting           &  10.3 & 21.8x &  28.5 &  7.9x &  55.4 & 4.1x \\ \hline
        Recursive binary splitting &   8.8 & 25.5x &  23.2 &  9.7x &  46.4 & 4.8x \\ \hline
        Purim                      &  14.6 & 15.4x &  31.1 &  7.2x &  55.2 & 4.1x \\ \hline
        Sobel-R1                   &   8.7 & 25.8x &  20.7 & 10.8x &  39.4 & 5.7x \\ \hline
    \end{tabular}
    \caption{\textbf{Expected time in days to test 10\% of the US population for three different infection rates.}
        Test capacities and simulation parameters as of April 14, 2020, see Table~\ref{tab:countries}
    }
    \label{tab:test10}
\end{table}
 
It is important to note that a screening campaign based on a hierarchical approach on average identifies fewer cases than individual testing. Assuming a test sensitivity of 0.99 for an individual test, employing hierarchical testing has a probability of 0.01 (or 1\%) to miss a certain case on each of several test stages. Fig~\ref{fig3}(C) shows that the (compounded) expected rate of identified cases (true positives) therefore drops by between 1\% and 5\% in total. Hence, the \textit{sensitivity} of a hierarchical approach is lower compared to individual testing. In contrast, the likelihood of incorrectly classifying a subject as infected is reduced from 1\% to almost 0\%, see Fig~\ref{fig3}(D); i.e. when carrying out a hierarchical approach the \textit{specificity} is improved compared to individual testing. 

To illustrate why the number of identified cases per test (ICPT) is a suitable measure to compare the effectiveness of different pooling methods, we simulate the application of 100,000 tests on a population of 1 million individuals, see Table~\ref{table3}.
We demonstrate that while the number of false negatives increases when employing pooling methods, two other measures change to a much greater degree: the number of untested (infected) individuals decreases \textit{and} the number of true positives increases.
As before, we assume an infection rate of 1\%. The effectiveness of each approach can directly be observed via the number of individuals tested in a given population. While for individual testing the number of tested individuals equals the number of tests available, 2-level pooling examines 490,792 individuals on average with 100,000 tests, and Sobel-R1 can test the whole population more than once (1.08 times). Binary splitting is best suited to illustrate the occurrence of false negatives: the overall sensitivity is reduced from 99\% to about 94.2\% (see Fig.~\ref{fig3}(C)), and the false negative rate increases correspondingly from 1\% to 6.8\%. In our simulation, the expected number of cases that are not correctly classified as infected increases from 10 to 471. However, binary splitting at the same time correctly identifies 8,363 more infected individuals which individual testing was not able to test. 
Our results clearly show that all pooling methods significantly outperform individual testing, with the additional benefit of fewer people being sent to quarantine despite being healthy (fp).

\begin{table}[!ht]
\centering

\resizebox{\textwidth}{!}{\begin{tabular}{|c+r|r|r|r|r|r|}
\hline
& \multicolumn{6}{|c|}{\bf Method}\\ 
& \multicolumn{1}{|p{0.1\linewidth}|}{Ind. testing} & \multicolumn{1}{|p{0.1\linewidth}|}{2-level \newline pooling} & \multicolumn{1}{|p{0.1\linewidth}|}{Binary \newline splitting} & \multicolumn{1}{|p{0.15\linewidth}|}{Recursive bin.~splitting} & \multicolumn{1}{|p{0.1\linewidth}|}{Purim} & \multicolumn{1}{|p{0.1\linewidth}|}{Sobel-R1} \\ \thickhline
\multicolumn{1}{|p{0.15\linewidth}+}{\centering No.~individuals~tested} & 100,000.0 & 490,791.9 & 804,985.4 & 976,797.2 & 720,729.5 & 1,083,181.0$^{(*)}$ \\ \hline
\multicolumn{1}{|p{0.15\linewidth}+}{\centering Cases found (tp)} &  990.0 & 4,810.7 & 7,578.5 & 9,353.0 & 6,993.6 & 10,442.2$^{(*)}$ \\ \hline
\multicolumn{1}{|p{0.16\linewidth}+}{\centering False~positives (fp)} & 990.3 & 460.5 & 76.6 & 459.5 & 393.6 & 850.4$^{(*)}$  \\ \hline
\multicolumn{1}{|p{0.15\linewidth}+}{\centering Cases missed (fn)} & 10.0 & 97.2 & 471.3 & 415.0 & 213.7 & 389.6$^{(*)}$ \\ \hline
\multicolumn{1}{|p{0.15\linewidth}+}{\centering Cleared (tn)} & 98,009.7 & 485,423.4 & 796,858.9 & 966,569.8 & 713,128.6 & 1,071,498.7$^{(*)}$ \\ \hline
\multicolumn{1}{|p{0.15\linewidth}+}{\centering Infected ind.~not even tested due to limited test capacity} & 9,000.0 & 5,092.1 & 1,950.1 & 232.0 & 2,792.7 & 
\parbox[t]{0.15\linewidth}{\raggedleft \hfill 0.0 \newline (-~831.8$^{(*)}$)} 
\\ \hline
\end{tabular}}
\caption{
{\bf Effectiveness of conducting 100,000 tests on a population of 1 million; full statistics over 10 simulations each.}
Simulating 100,000 tests on a total population of 1 million with an infection rate of 1\%. We assume a sensitivity of 99\% and a false negative rate of 1\% for a single PCR, and no dilution due to pooling. We report how many individuals are tested with 100,000 (pooled) tests as well as true positives (tp), false positives (fp), false negatives (fn) and true negatives (tn). The reported numbers are averaged over 10 runs testing the whole population, measuring a relative standard deviation of $<0.3\%$ each. (*) Note that Sobel-R1 is able to test the whole population 1.08 times; this methods identifies $10,442.2 > 10,000$ cases, identifying 832 cases in the second run leading to a negative value in the last entry.
}
\label{table3}
\end{table}

Note that with binary splitting a single sample can undergo up to six sequential testing steps. Purim and 2-level testing can be carried out in two sequential testing steps. For recursive binary splitting and Sobel-R1, re-pooling the batch sizes can lead to large numbers of sequential stages (up to 17 and 23 hierarchical steps in the worst case scenario for multiple cases in a pool, respectively). For Sobel-R1 as the (theoretical) optimal method, at an infection rate of $1\%$, only 5\% of tests could be carried out with at most 5 hierarchical steps, whereas 95\% could be carried out with at most 13 hierarchical steps.

\section{Discussion}

To the best of our knowledge, this study presents the first comprehensive, comparative assessment of optimised testing strategies for effective large-scale screening for infection with SARS-CoV-2. Our simulations indicate that population-level diagnostics in a pandemic the scale of COVID-19 will only be possible by making use of pool-based strategies -- at the time of our study in May 2020, testing even 10\% of the US population with the available testing capacity would have take more than seven months. 

Our study enables the following key conclusions for mass testing for COVID-19:

\begin{enumerate}
     \item Screening the entire population can be carried out several times faster via hierarchical or matrix-based approaches compared with individual testing in every considered scenario.
    \item Among the methods that can be considered immediately applicable in practice, we see that with respect to an optimal ICPT:
    \begin{enumerate}
        \item Binary splitting is the best method for infection rates between 0\% and $2.5$\%.
        \item Purim is the best method for infection rates between $2.5$\% and 12\%.
        \item 2-level pooling is the best method for infection rates beyond 12\%.
    \end{enumerate}
    \item Recursive binary splitting and the Sobel-R1 method would allow to significantly improve the ICPT. However, to be practically applicable, first, either the time required to process each individual test would have to be reduced or the possible storage time of samples would have to be increased by a factor of about two (given the average number of hierarchical steps expected). Second, sufficient sample material would have to be available for a large number of repeat tests. Third, software support for lab technicians would need to be provided to guide lab operations and ensure that the method can be carried out swiftly and correctly.
\end{enumerate}

With a maximum pool size of 32, binary splitting requires up to six sequential testing steps for a single sample. With an assumed duration of 5 hours per test, individual results are available after $30$h and six subdivisions of the sample for a pool size of 32, which we consider as a realistic scenario. Purim and 2-level testing can be carried out in two sequential testing steps. These methods conclude testing after $10$h and can realistically be applied in practice as of now.

Of note, all hierarchical models require adequate scheduling of the processes within the involved laboratories, as the performance of certain tests is conditional upon the outcome of previous tests. However, these scheduling problems are well studied, with standard-type optimisations available that can easily be implemented.

For the purpose of this study, algorithms were implemented as described in the literature. Therefore, certain adaptations to account for medical and legal requirements might be needed (e.g. Sobel-R1 can identify cases via exclusion, requiring an additional confirmation step to be included for full clinical applicability).

In recent publications and preprints published in the course of the COVID-19 pandemic, 2-level pooling with optimised pool size, matrix-based approaches for pooled screening and binary splitting were presented.\cite{AustrianPaper,PURIM, Taufer2020.04.06.028431, Sinnott-Armstrong2020.03.27.20043968, Shani-Narkiss2020.04.06.20052159, Noriega2020.04.03.024216, Narayanan2020.04.03.20051995} Our simulations include models of all of these central categories of approaches, benchmarking and comparing the various testing strategies, and demonstrating for which scenarios each method can become the optimal choice.
Our model calculations prioritise identifying infected individuals as soon as possible to prevent further spread, i.e.\ optimising for speed of completion of a screening campaign. This is especially important since regular re-testing will be necessary given the continuous risk of infection and potentially even recurrence as highlighted by recent case reports.

Of course, differences in test characteristics -- sensitivity, rate of false positives, and processing time -- have a significant impact on the choice of testing strategy. In particular, test sensitivities were recently reported to be in the range of 0.75 for sputum or nasal swabs.\cite{Yang2020.02.11.20021493} However, based on the same study, the decreased sensitivity can mostly be attributed to the quality of the samples and variations in distribution of viral load in patients -- in other words, the samples taken will or will not contain viral material that could be amplified during the PCR tests (with their sensitivity remaining constant at 0.99). Thus, if a sample contains no viral material, then any method will fail, whether individual or pooled. If a sample, however, contains viral material, pooling -- which happens after sample taking -- can be conducted and is advisable as discussed in this paper. For illustration, a 0.75 test sensitivity scenario is given in the supplementaries in~\ref{S1_Fig} and~\ref{S2_Fig}, showing that even under these assumptions, pooling is still advisable. 

In addition, all variations can be modelled via our online tool 
\begin{quote}
	\url{https://covid19.enfunction.com}.
\end{quote}

We show that a hierarchical testing approach increases the specificity of a screening campaign. The reduced number of false positives is important, as it equates to ca.\ 1,000 healthy individuals who would otherwise have been erroneously quarantined in a population of 100,000. At the same time, binary and recursive binary splitting correctly identify up to ten times more cases per test compared to the currently employed individual testing. Furthermore, they enable multiple screening campaigns within the same time normally required for a single screening campaign where each sample is tested individually. 

One of the biggest caveats of any modelling approach is the need to show that theoretical simulations can be successfully translated into public health measures. Hogan et al.\ and Yelin et al.\ recently presented their findings on the practical applicability of sample pooling in California and Israel, respectively, thereby providing important experimental validation for our modelling approach.\cite{10.1001/jama.2020.5445, IsraelPaper} In fact, Yelin et al.\ even showed that a pool size of up to 64 still provides an acceptable sensitivity, potentially enabling larger pool sizes that would increase the benefits of hierarchical pooling approaches. The exact maximum number of samples that can be pooled and analysed while maintaining adequate sensitivity will likely depend on the characteristics of each test and is an important question for follow-on work. Of note, on 13 April 2020, the Indian Council of Medical Research published guidance recommending limited 2-level pooling to increase screening capacity -- to our best knowledge the first pooling approach formally adopted as of now, but certainly not the last.\cite{icmr}

\section{Conclusion}

Rapid identification of patients, asymptomatic carriers, and the modes of transmission of a given pathogen are key goals of pandemic response, that can then be embedded into a larger set of medical countermeasures.\cite{simpsonlancet} This study provides a theoretical framework and practical guidance to frontline medical staff, public health authorities, and governments on how to best deploy limited testing resources to maximise the number of people tested in the shortest amount of time possible -- in the case of COVID-19 as well as future pandemic outbreaks.

\section{Acknowledgements}
The authors would like to thank Professor Roy Kishony (Technion - Israel Institute of Technology) for valuable contributions on PCR test characteristics; Dr Shmona Simpson (Bill and Melinda Gates Foundation, USA) for valuable discussions regarding planning and implementation of pandemic response measures; Dr Stefan Zimmer (University of Stuttgart, Germany) for assistance in code review and discussions regarding optimised pooling approaches; Dr Nikita Kaushal (Nanyang Technological University, Singapore) for manuscript review; and Dr Semen Trygubenko (Arctoris, UK) for assistance in implementing the online tool that allows for user-directed modelling for different testing scenarios.

\section{Funding Statement} 
TdW, DP and MIB are supported by the Young Academy of the German National Academy of Sciences. TdW is funded by the German Research Foundation (DFG) grant WO 2206/1-1 under the Emmy Noether Programme. DP and MR are funded by the Cluster of Excellence Data Integrated Simulation Science under Germany’s Excellence Strategy (EXC 2075 - DFG grant 390740016). Arctoris Ltd provided support in the form of salary for MIB. The specific roles of these authors are articulated in the ‘author contributions’ section. The authors acted independently and the authors' funding sources did not have any involvement in the study design, data collection and analysis, decision to publish, or preparation of the manuscript.

\section{Author contributions}
Data curation: TdW, DP, MR, JH, MIB.
Funding acquisition: TdW, DP, MIB.
Investigation: TdW, DP, MR, JH, MIB.
Methodology: TdW, DP, MR, JH, MIB.
Project administration: TdW, DP, MIB.
Supervision: TdW, DP, MIB.
Validation: TdW, DP, MIB.
Visualization: TdW, DP, MR, JH, MIB.
Writing – original draft: TdW, DP, MIB.
Writing – review \& editing: TdW, DP, MR, JH, MIB.

\section{Declaration of interests} 
MIB is a shareholder and director of Arctoris Ltd. This does not alter our adherence to PLOS ONE policies on sharing data and materials. Apart from salary support, Arctoris Ltd had no additional role in the study design, data collection and analysis, decision to publish, or preparation of the manuscript. 

\bibliographystyle{vancouver}
\bibliography{main_arXiv}

\addresseshere

\clearpage

\section{Supplementaries}

The following Figures~\ref{S1_Fig} and~\ref{S2_Fig} complement our study with plots for a decreased test sensitivity of $p=0.75$.
\setcounter{figure}{0}
\makeatletter 
\renewcommand{\thefigure}{S\@arabic\c@figure}
\makeatother
\begin {figure}[h]
\vspace{2em}
    \resizebox{\linewidth}{!}{
    \includegraphics[width=.02\linewidth]{./groupsize_legend.png}
    \includegraphics[width=.3\linewidth]{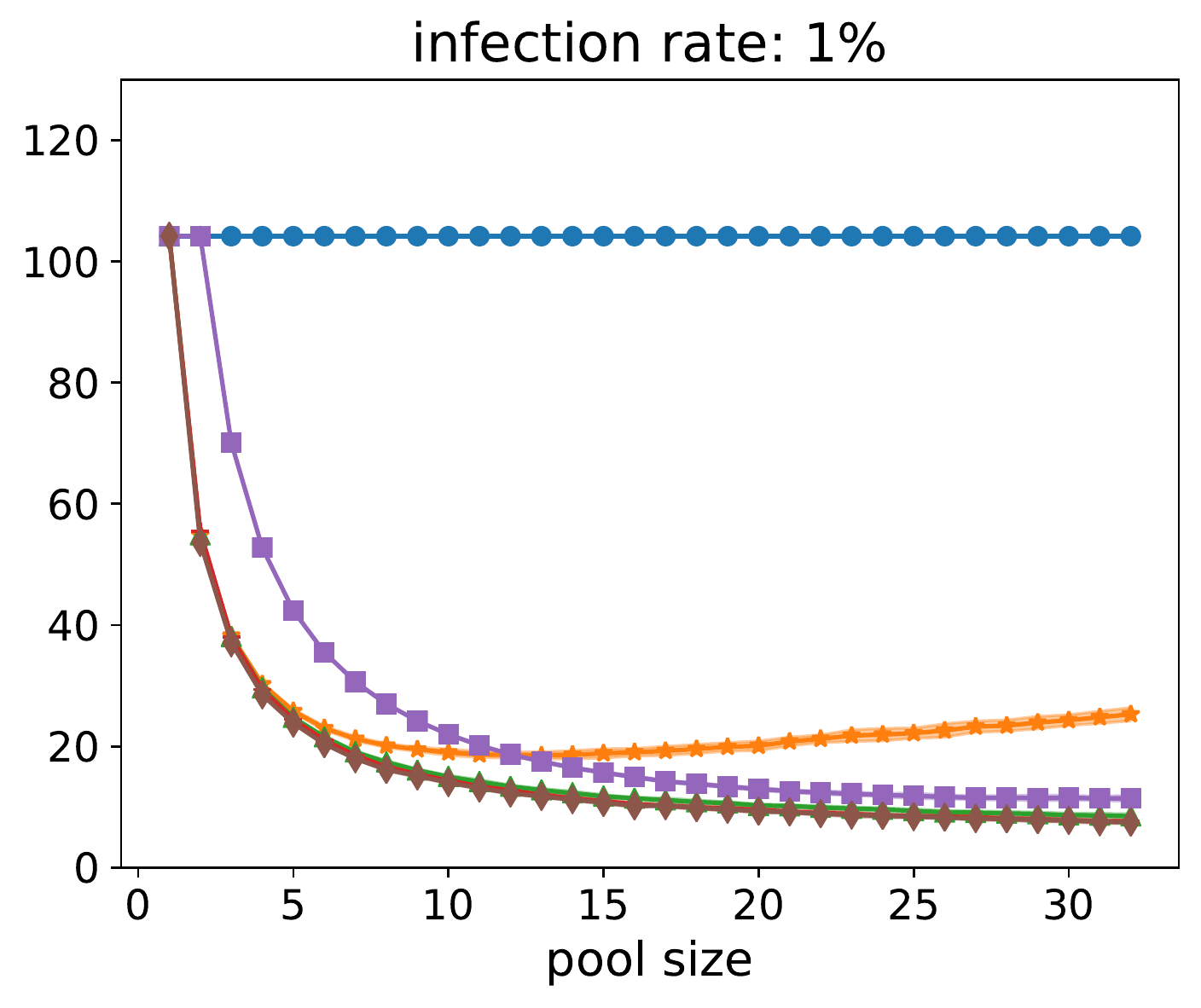}
    \includegraphics[width=.3\linewidth]{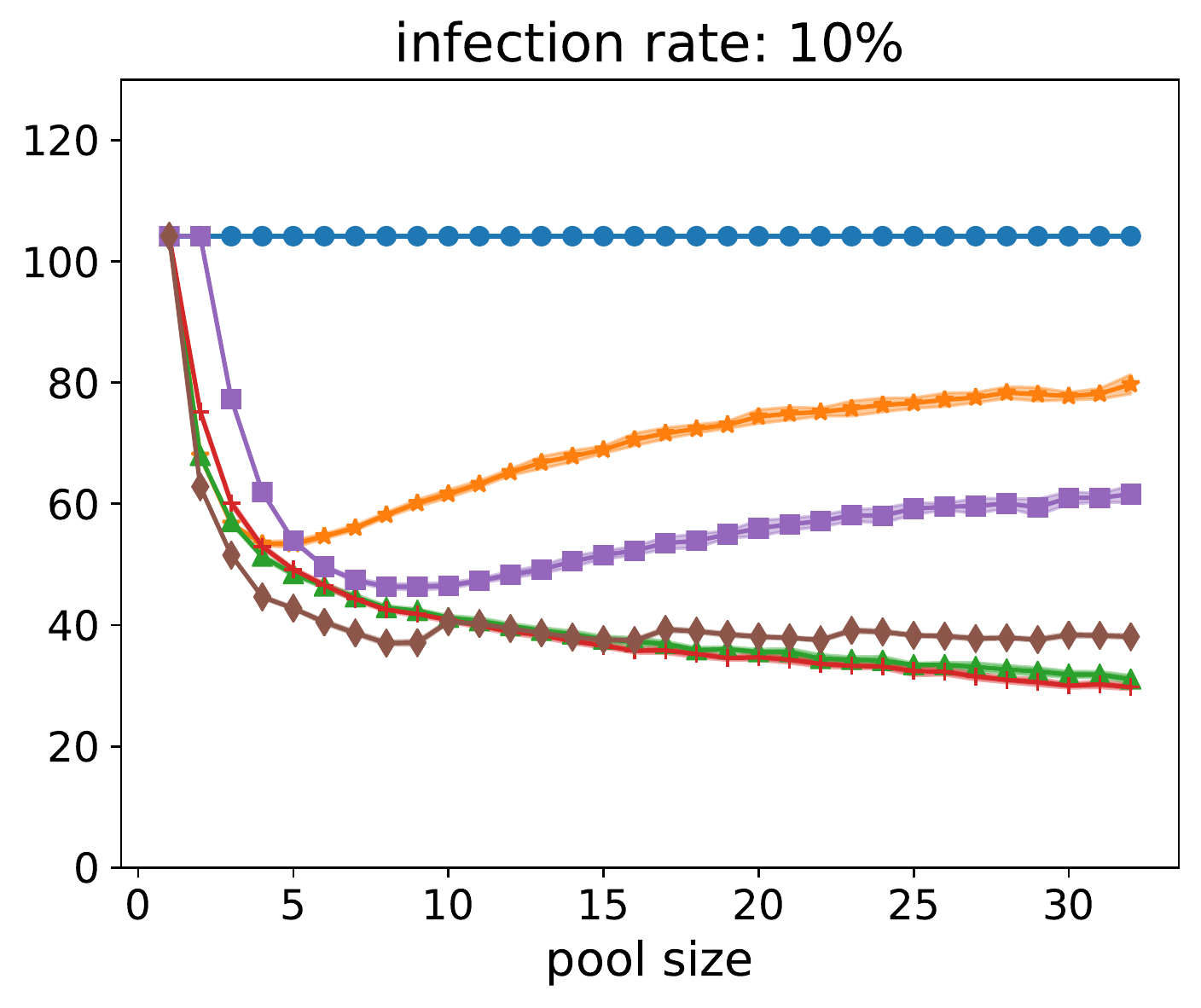}
    \includegraphics[width=.3\linewidth]{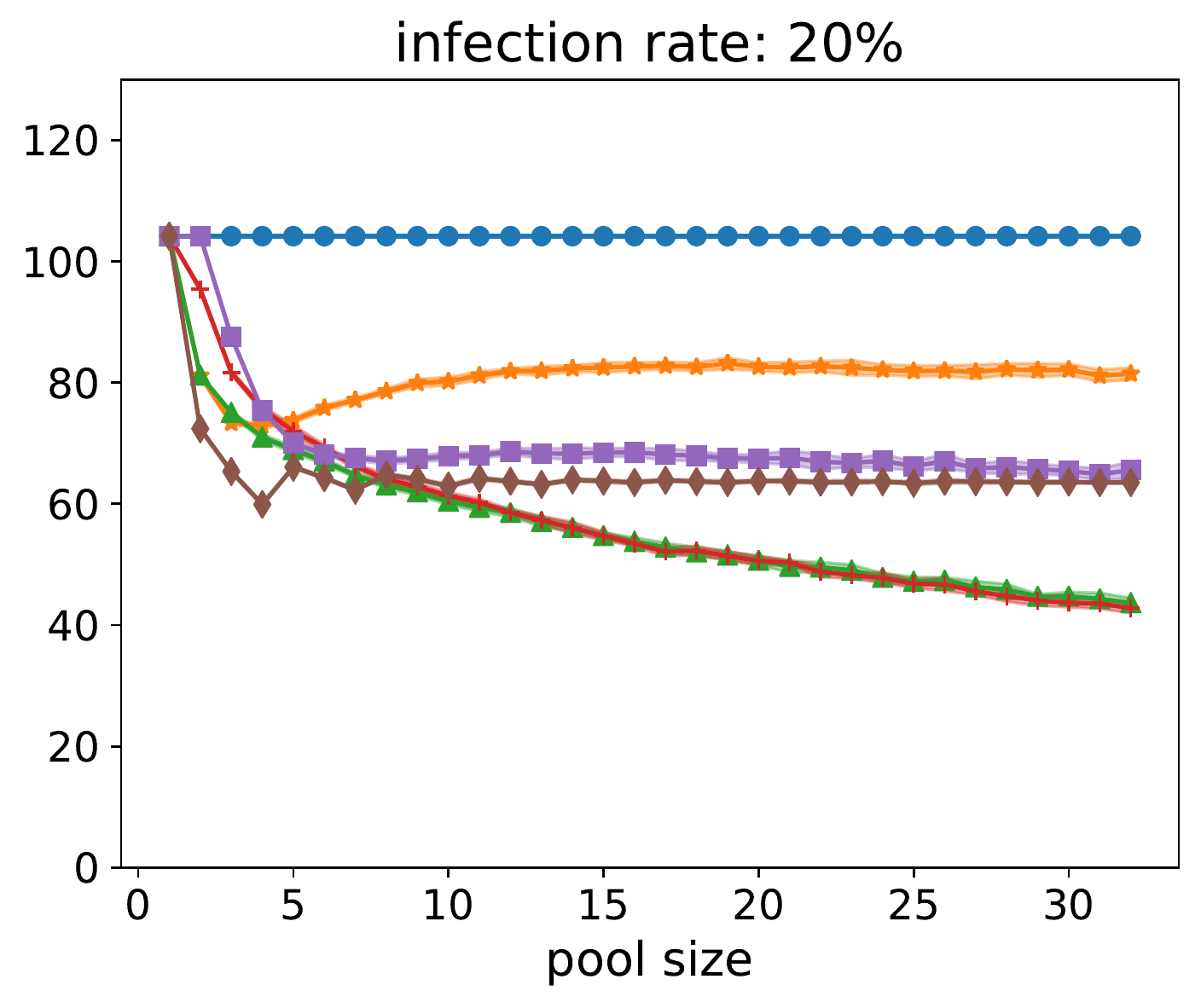}
    }
    \caption{The best pool size (lowest total time) depends on the infection rate, here for $ir$=1\%, 10\%, 20\%. Parameters: sensitivity $p=0.75$, false positive rate $q=0.01$, population $50,000$, test duration 5h, averaged over 10 runs. Blue: individual testing; orange: 2-level pooling; green: binary splitting; red: recursive binary splitting; purple: Purim; brown: Sobel-R1}
    \label{S1_Fig}
\end{figure}

\begin{figure}[h]
\centering
\begin{subfigure}[t]{\linewidth}
\centering
\includegraphics[width=0.45\linewidth]{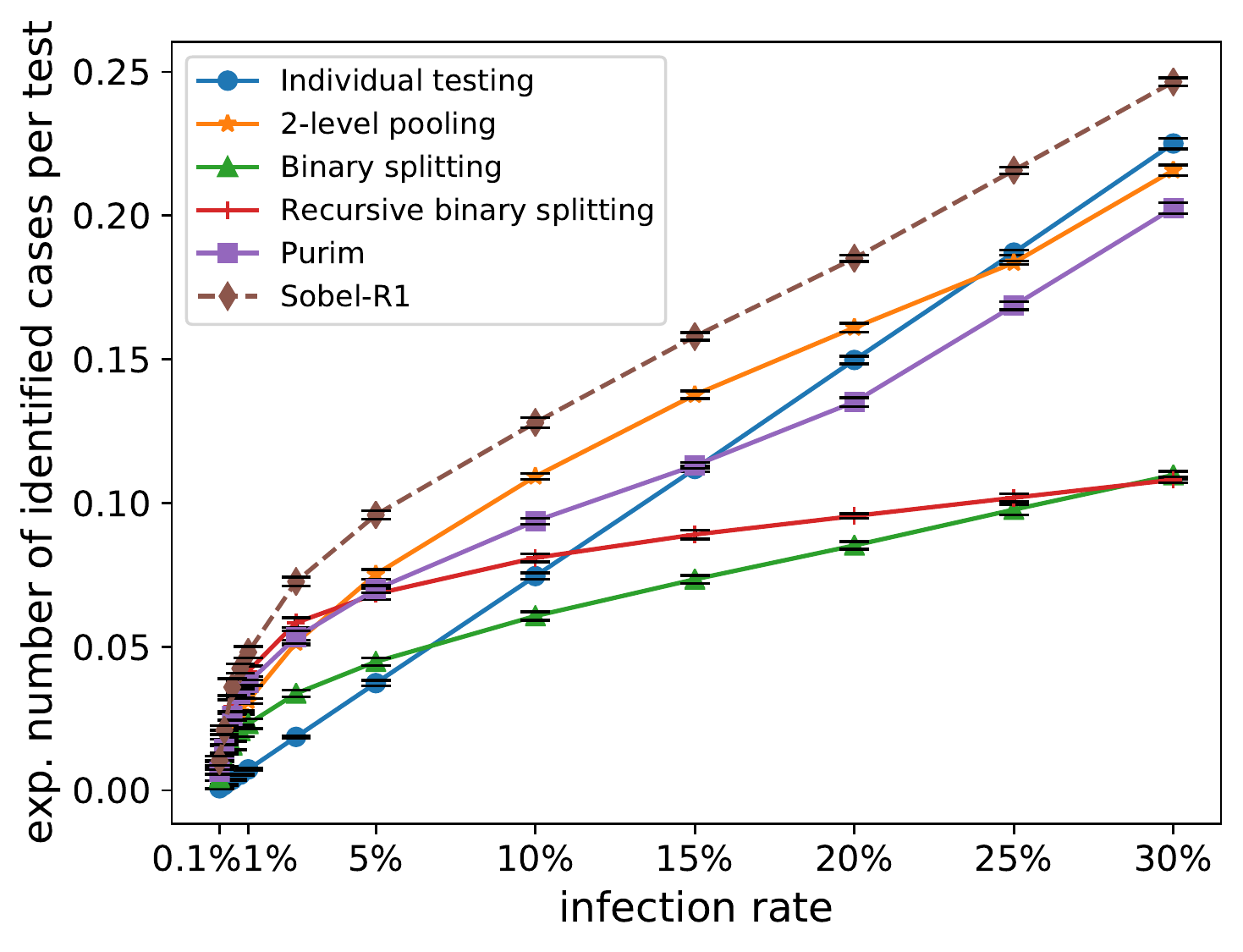}
    \hspace{0.02\linewidth}
    \includegraphics[width=0.45\linewidth]{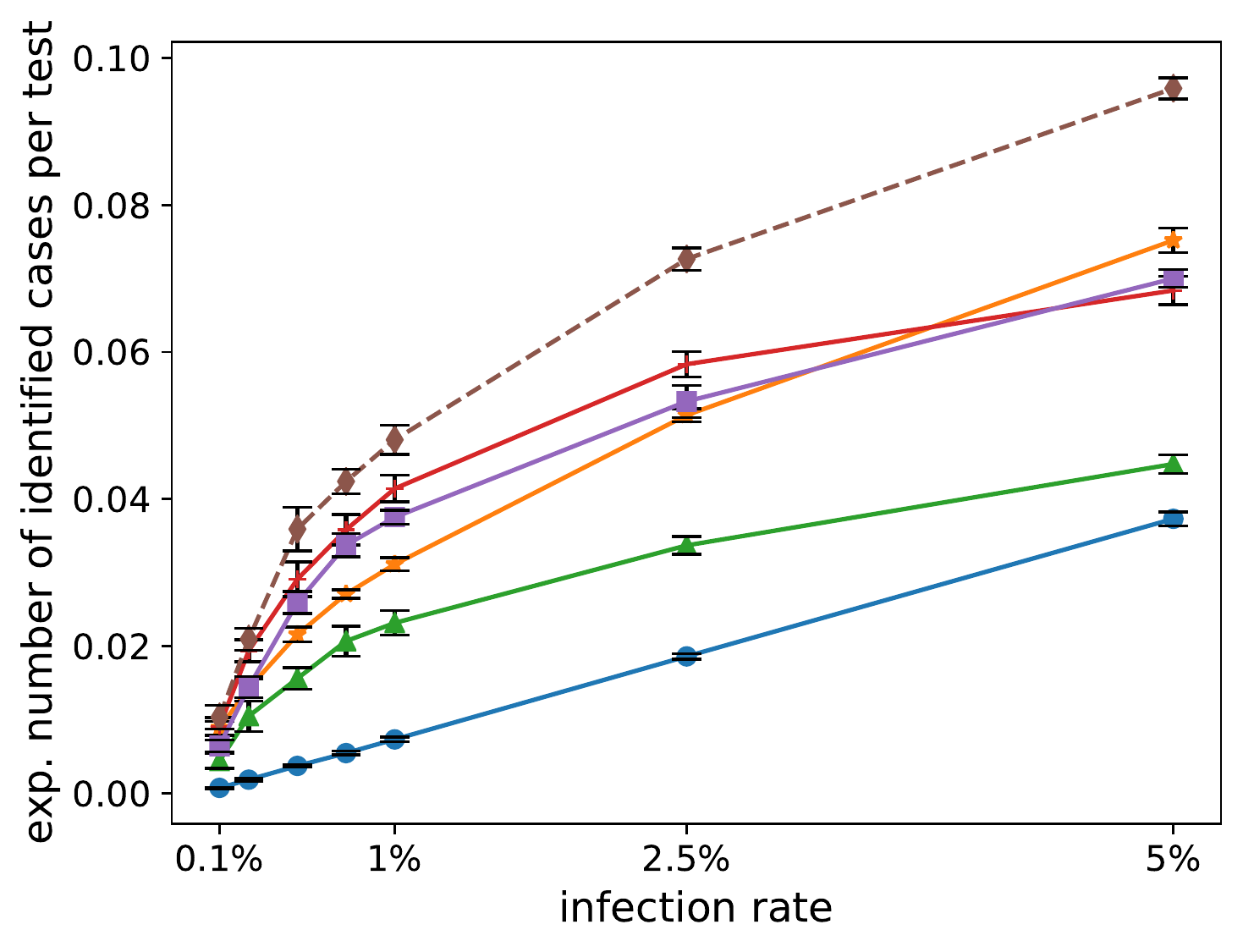}
\caption{Expected number and standard deviation of identified cases per test for different infection rates with optimised group sizes; right: zoom-in for small infection rates} 
    \label{fig:icpt075}
\end{subfigure}
%
\begin{subfigure}[t]{\linewidth}
\centering
    \includegraphics[width=0.45\linewidth]{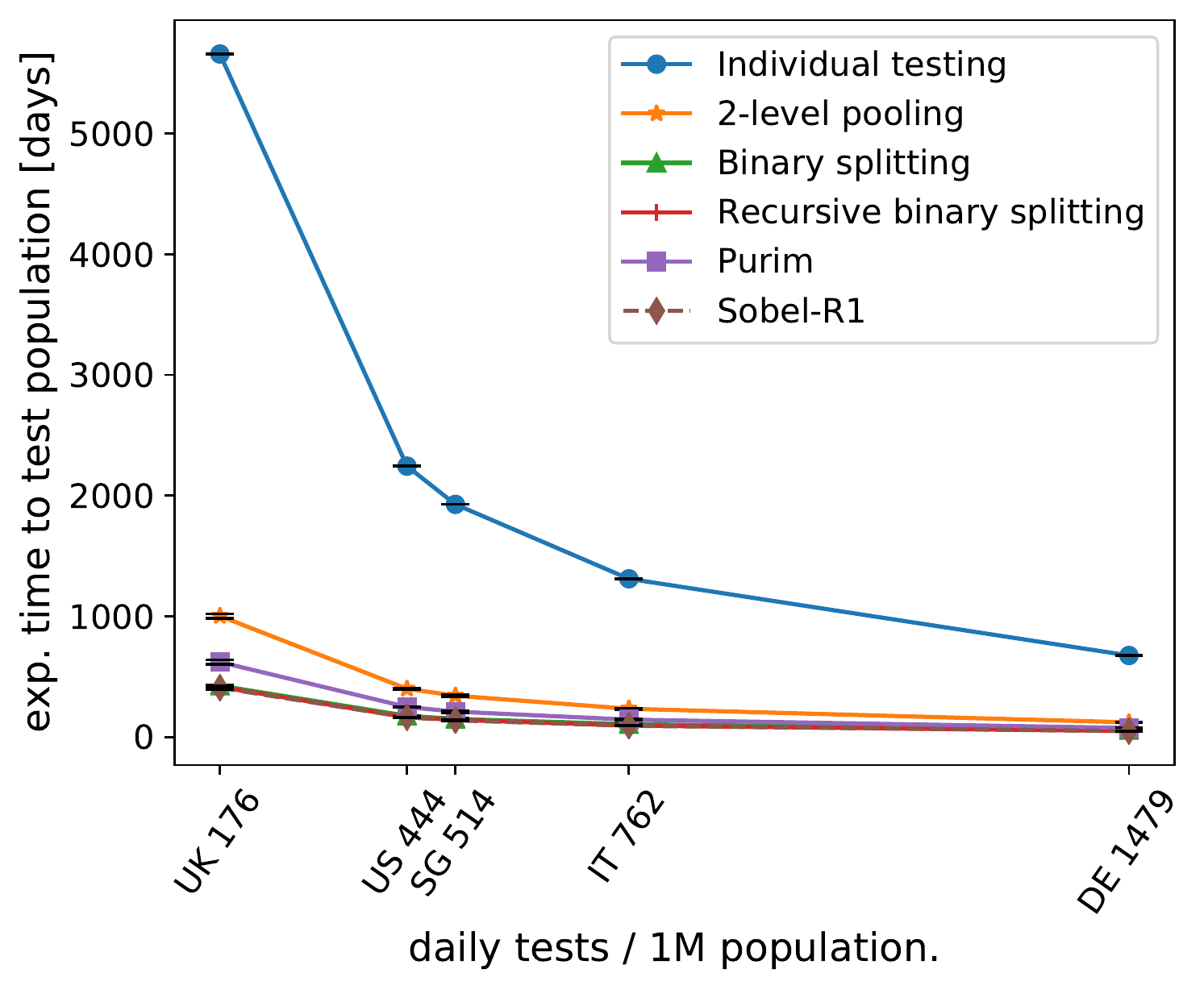}
        \hspace{0.02\linewidth}
    \includegraphics[width=0.45\linewidth]{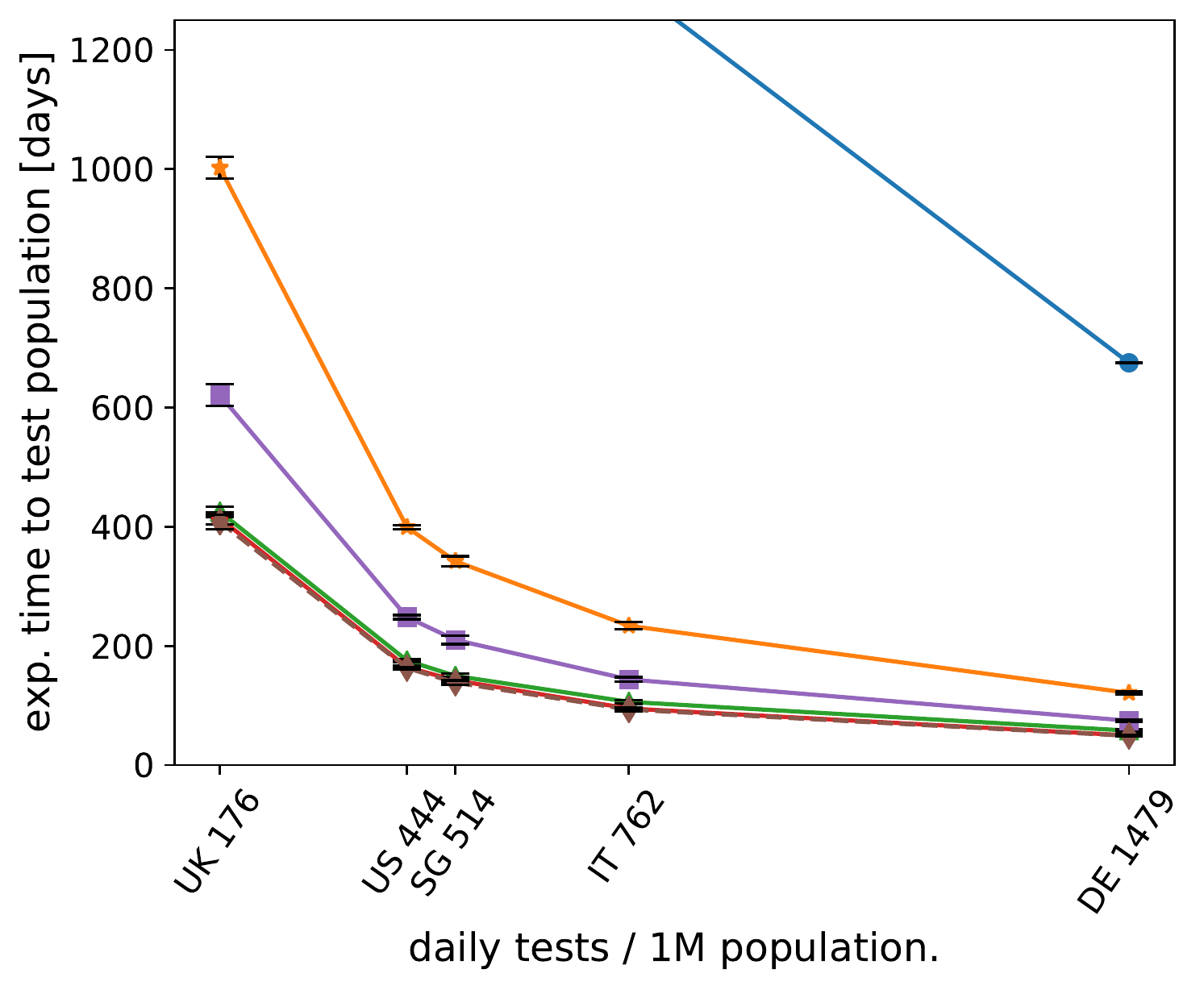}
    \caption{Expected time in days and standard deviation to test the whole population depending on daily test capacity per 1m population, infection rate 1\%; right: zoom-in excl.\ individual testing}
    \label{fig:time_test075}
\end{subfigure}
\begin{subfigure}[t]{.45\linewidth}
    \includegraphics[width=\linewidth]{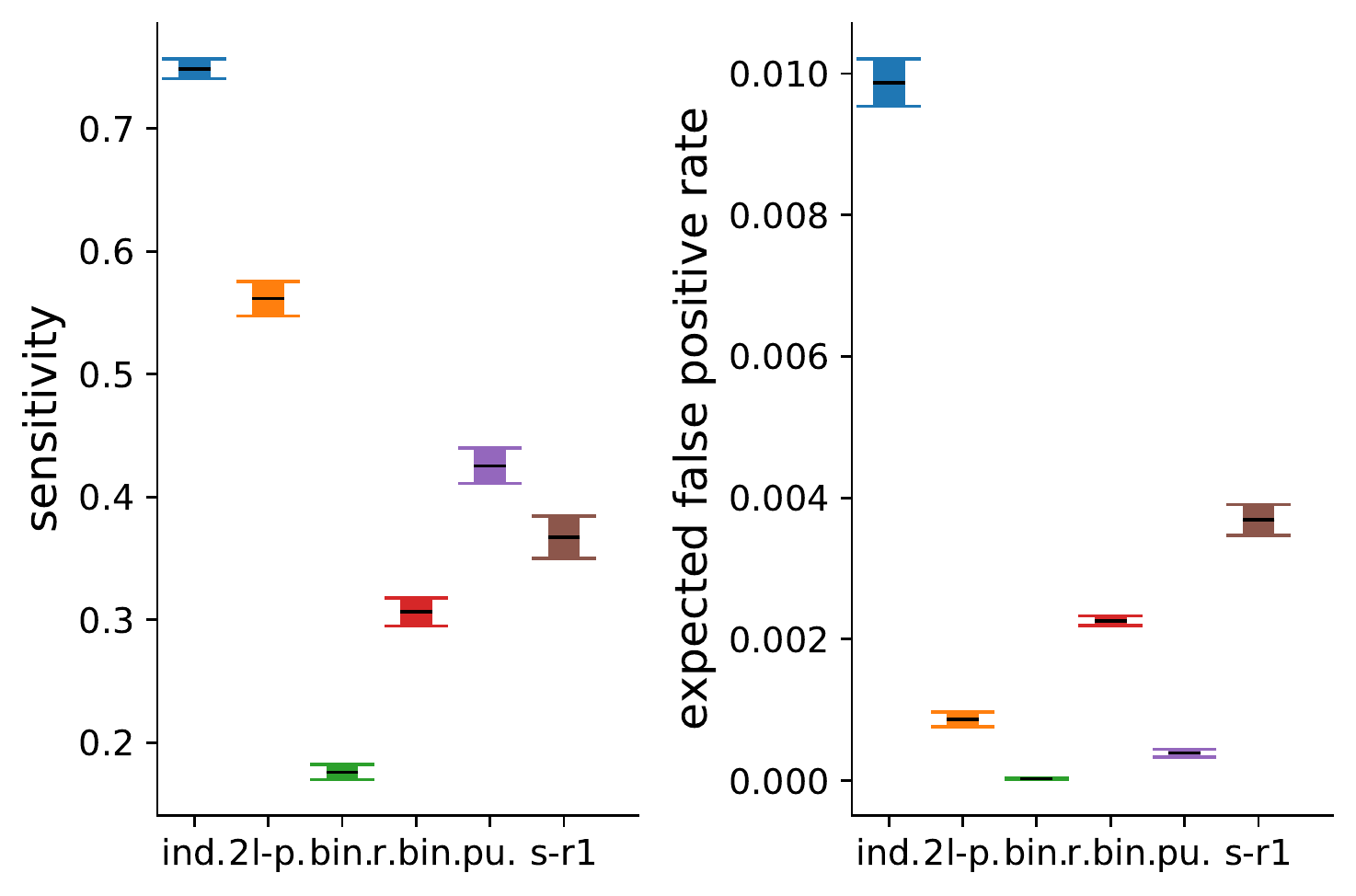}
    \caption{Expected true positives and false positives for the six methods (pop.~100,000)}
    \label{fig:tp_fn075} 
\end{subfigure}
\hspace{0.02\linewidth}
\begin{subfigure}[t]{.45\linewidth}
    \includegraphics[width=\linewidth]{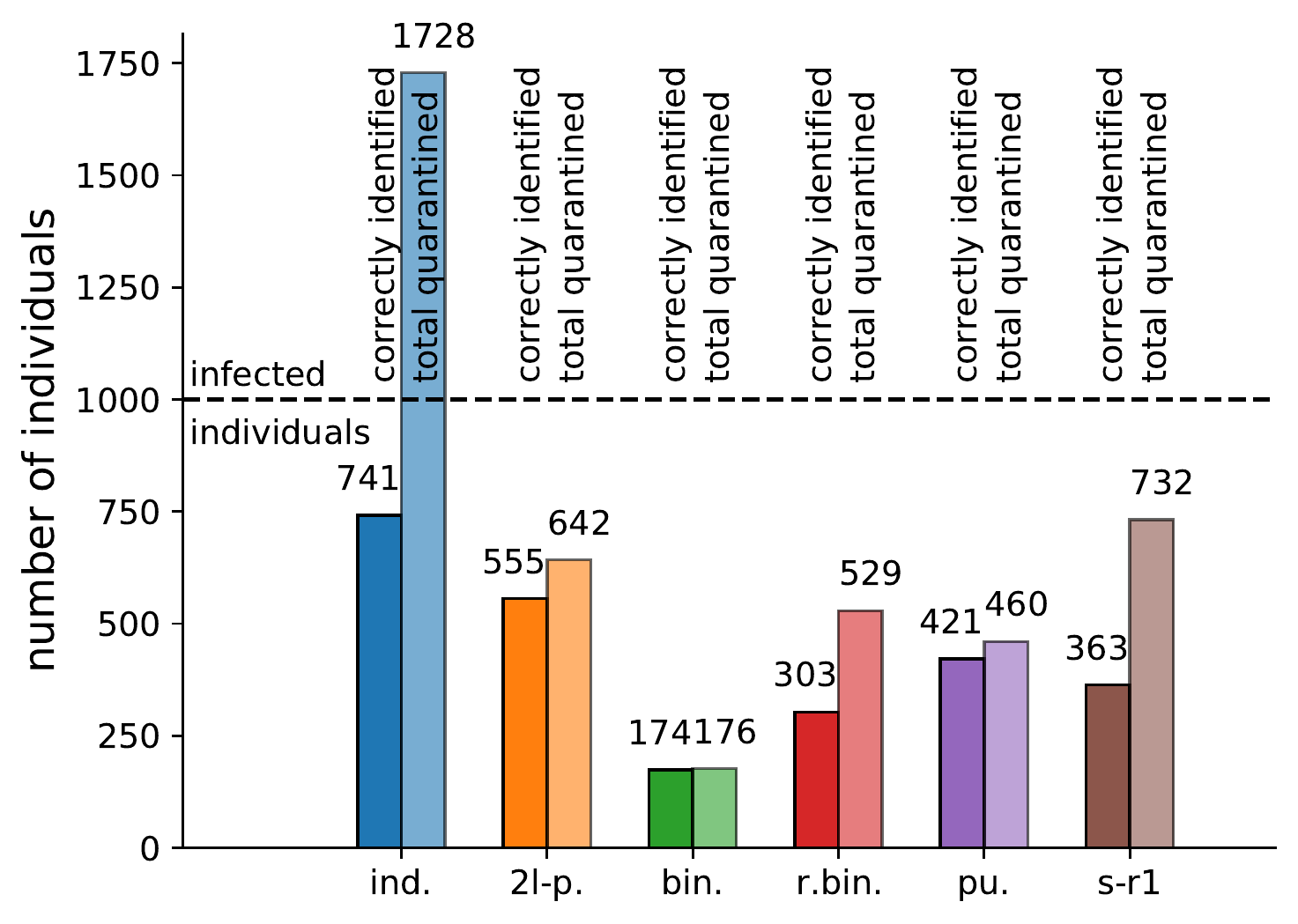}
    \caption{Numbers of identified cases and quarantined individuals (pop.~100,000)}
    \label{fig:quarantined075}
\end{subfigure}
    \label{fig:scenario1075}
\caption{Screening the whole population. 
Parameters: sensitivity $p=0.75$, false positive rate $q=0.01$, test duration 5h, averaged over 10 runs. Optimal (max.) pool size each (c.f.~Fig~\ref{fig2}); for $ir$=1\% as in (\ref{fig:time_test075})--(\ref{fig:quarantined075}) we obtain individual testing: 1; 2-level pooling: 12; binary splitting: 32; recursive binary splitting: 32; Purim: 31, Sobel-R1: 32}
 \label{S2_Fig}
\end{figure}

\end{document}